\documentstyle[psfig]{mn}
\def\etal{{\rm et al.}}

\def\P3M{P$^3$M}
\def\vec#1{{\bf #1}}

\def\and{, }
\font\mgn=cmti7
\def\eqnote#1{\marginpar{\mgn #1}}
\def\eqnote#1{{}}

\newfam\cyrfam
\font\tencyr=wncyr10 at 12pt
\font\sevencyr=wncyr7 at 8.5pt \font\fivecyr=wncyr5 at 6pt

\textfont\cyrfam=\tencyr \scriptfont\cyrfam=\sevencyr
            \scriptscriptfont\cyrfam=\fivecyr

\def\fracnum#1#2{\raise 2.1pt\hbox{$\scriptstyle #1$}\kern
-1.2pt/\kern -1.2pt \lower 2.1pt\hbox{$\scriptstyle#2$}\,}

\begin{document}

\title[Gravitational lensing] {Weak gravitational lensing in different cosmologies, using an algorithm for shear in three dimensions.} 

\author[Andrew J. Barber \etal] {Andrew J. Barber$^1$\thanks{Email: 
abarber@star.cpes.susx.ac.uk}, Peter A. Thomas$^1$, H. M. P. Couchman$^2$ and
C. J. Fluke$^3$ \\
{}$^1$Astronomy Centre, University of Sussex, Falmer, Brighton, BN1 9QJ\\ 
{}$^2$Department of Physics and Astronomy, McMaster University, Hamilton,
Ontario, L8S 4M1, Canada\\
{}$^3$Centre for Astrophysics and Supercomputing, Swinburne University, Melbourne, Australia
}

\date{Accepted 1999 ---. Received 1999 ---; in original form 1999 ---}

\maketitle

\begin{abstract}

We present the results of weak gravitational lensing statistics in
four different cosmological $N$-body simulations. The data has been
generated using an algorithm for the three-dimensional shear, which
makes use of a variable softening facility for the $N$-body particle
masses, and enables a physical interpretation for the large-scale
structure to be made. Working in three-dimensions also allows the
correct use of the appropriate angular diameter distances.

Our results are presented on the basis of the filled beam
approximation in view of the variable particle softening scheme in our
algorithm. The importance of the smoothness of matter in the universe
for the weak lensing results is discussed in some detail.

The low density cosmology with a cosmological constant appears to give
the broadest distributions for all the statistics computed for sources
at high redshifts. In particular, the range in magnification values
for this cosmology has implications for the determination of the
cosmological parameters from high-redshift Type Ia Supernov\ae. The
possibility of determining the density parameter from the
non-Gaussianity in the probability distribution for the convergence is
discussed.

\end{abstract}


\section{INTRODUCTION}
 
\subsection{Outline}

We present the results of a study of the weak gravitational lensing of
light in four different cosmological models, using the algorithm for
the three-dimensional shear developed by Couchman, Barber and Thomas
(1999). Since weak lensing effects depend on the angular diameter
distances for the lenses and sources, and also the specific
distribution and evolution of matter, the results are sensitive to the
particular cosmological model.

In {\em strong} lensing studies, frequent use is made of the
`thin-screen approximation,' in which the mass distribution of the
lens is projected along the line of sight and replaced by a mass sheet
with the appropriate surface density profile. Deflections of the light
from the source are then considered to take place only within the
plane of the mass sheet, making computations for the light deflections
simple.  The simplicity of the thin-screen approximation has also lead
to its frequent use in weak gravitational lensing studies, where each
of the output volumes from cosmological $N$-body simulations is
treated as a planar projection of the particle distribution within it.

However, a number of problems can arise with two-dimensional
approaches, especially in weak lensing studies in which the
large-scale distribution of matter extending to high redshifts is
responsible for the lensing. Couchman, Barber and Thomas (1999)
considered some of the shortcomings and were motivated to develop an
algorithm to compute the six independent components in three
dimensions of the second derivative of the gravitational potential
(the three-dimensional shear). In the method used in the present work,
for which the effects of lensing along lines of sight are required,
the two-dimensional `effective lensing potentials' (see Section 2) are
obtained by integrating the computed three-dimensional shear
components along the lines of sight. These effective lensing
potentials are used to construct the Jacobian matrices which are
recursively generated along the lines of sight, and from the final
Jacobian matrices the magnifications and two-dimensional shear are
determined.

A brief outline of this paper is as follows.

In {\bf Section 1.2} we summarise the previous weak gravitational
lensing methods by other authors.

In {\bf Section 2}, the essential equations for gravitational lensing
are explained, and the multiple lens-plane theory is described,
culminating in expressions for the magnification, convergence,
two-dimensional shear, and ellipticity, which are the weak lensing
outputs from the cosmological simulations.

{\bf Section 3.1} summarises the three-dimensional shear algorithm,
and particularly the choice of the variable softening scale for the
particles. This feature limits the effects of isolated particles (not
representative of the large-scale structure), and builds in a physical
significance both to the choice of the softening scale, and to the
underlying form of the dark matter. We explain also how the softening
carefully limits the incidence of strong lensing events in all the
cosmologies. In {\bf Section 3.2} we describe the four cosmological
$N$-body simulations used in this work. In {\bf Section 3.3} we state
how the simulation boxes are oriented to minimise correlations in the
large-scale structure between adjacent time-outputs, and we describe
the establishment of lines of sight through the simulations, and the
locations of the evaluation positions for the shear within them. In
{\bf Section 3.4} we explain how the computed shear values are
converted (after integration along the lines of sight) to physical
units, and the approximations we have adopted in the method.

In {\bf Section 4.1} the Dyer-Roeder equation for the angular diameter
distances is introduced and {\bf Section 4.2} summarises the results
derived in the Appendix for the beam equation generalised for all the
cosmologies. This precedes the discussion ({\bf Section 4.3}) of
magnifications obtainable in inhomogeneous universes with different
degrees of smoothness, and highlights the significance of differences
between researchers who adopt different approaches to this subject.

{\bf Section 5} formally presents the weak lensing results. We first
({\bf Section 5.1}) attempt to see if the onset of structure formation
can be seen from the shear data, and briefly comment on the expected
behaviour of the developing shear. Secondly ({\bf Section 5.2}), the
results for magnification, convergence, shear and ellipticity are
presented for all the cosmologies, compared and contrasted. We
consider the impact of the smoothness parameter on the results.

It may be possible to determine the density parameter, $\Omega_0$,
from the probability distribution and the skewness in the distribution
for the convergence, both of which may be measurable
observationally. In {\bf Section 6.1} we present our results for these
quantities, and compare them with others. The application to the
determination of the cosmological parameters from Type Ia Supernov\ae~
is mentioned in {\bf Section 6.2}.

In {\bf Section 7} the weak lensing statistics detailed in Section 5
are summarised, together with the results of the non-Gaussianity in
the convergence. We also compare and contrast our results with other
authors. Of particular significance are the common use of
two-dimensional approaches by others, and the use of either point mass
particles or particles with small softening scales, which introduce
high values of magnification. These authors often use the empty cone
approximation, rather than the full beam approximation, making
comparisons difficult. This latter point is discussed.

In the {\bf Appendix} we state the generalised beam equation, and
derive the equations necessary for the numerical determination of the
angular diameter distances for all the cosmologies.

\subsection{Previous work}

Numerous methods have been employed to study weak lensing, and
throughout this paper we will make comparisons with previous work by
other authors. We summarise here the methods which have been used by
others. 

Jaroszy\'{n}ski et al. (1990) use a `ray-tracing' method to evaluate
the matter column density in a matrix of 128$^3$ pixels for each of
their lens-planes. The boxes were generated using a particle-mesh (PM)
code in the standard Cold Dark Matter (SCDM) cosmology, and were of
side-dimension 128$h^{-1}$Mpc, where $h$ is the Hubble constant in
units of 100~km~s$^{-1}$~Mpc$^{-1}$. By making use of the assumed
periodicity in the particle distribution orthogonal to the line of
sight, they translate the planes for each ray, so that it becomes
centralised within the plane of one full period in extent. This
removes any bias acting on the ray when the shear is computed. Instead
of calculating the effect of every particle on the rays, the pixel
column densities in the single period plane are used, and they assume
that the matter in each of the pixels resides at the centre point of
each pixel. They calculate the two two-dimensional components of the
shear (see Section 2 for the definition of shear) as ratios of the
mean convergence of the beam, which they obtain from the mean column
density. However, they have not employed the net zero mean density
requirement in the planes, (described in detail by Couchman, Barber
and Thomas, 1999), which ensures that deflections and shear can only
occur when there are departures from homogeneity. Also, the matter in
the pixel through which the ray is located is excluded. To follow the
shearing across subsequent planes they recursively generate the
developing Jacobian matrix for each ray, in accordance with the
multiple lens-plane theory (see Section 2).

Wambsganss, Cen and Ostriker (1998) also use the `ray-tracing' method in
cosmological $N$-body simulations. The method is applied to PM
simulations of the SCDM cosmology, in which a convolution method is
used to combine large-scale boxes of 400$h^{-1}$Mpc and resolution
0.8$h^{-1}$Mpc, with small-scale boxes of 5$h^{-1}$Mpc and the higher
resolution of 10$h^{-1}$kpc. They randomly orient each simulation box,
and project the matter contained within each onto a plane divided into
pixels. They choose the central $8h^{-1}$Mpc $\times~8h^{-1}$Mpc
region through which to shoot rays, but account for the deflections of
the rays in terms of all the matter in the plane of $80h^{-1}$Mpc
$\times ~80h^{-1}$Mpc. However, to speed up the computation, a
hierarchical tree code in two dimensions is used to collect together
those lenses (pixels) far away, whilst treating nearby lenses
individually. The code assumes that all the matter in a pixel is
located at its centre of mass. The matter in each pixel, which
measures $10h^{-1}$kpc $\times~10h^{-1}$kpc, is assumed to be
uniformly spread.  By using the multiple lens-plane theory, they show
both the differential magnification probability distribution, and the
integrated one for 100 different source positions at redshift
$z_s=3.0$. One advantage of this type of ray-tracing procedure is its
ability to indicate the possibility of multiple imaging, where
different rays in the image plane can be traced back to the same pixel
in the source plane, and they are able to compute the statistics of
angular separations for multiple images.

Marri and Ferrara (1998) use lens-planes up to redshifts of $z=10$ for
mass distributions determined by the Press-Schechter formalism, and
treat the particles as point-like masses with no softening. They apply
their ray-tracing method to three cosmologies with
$(\Omega_M,~\Omega_{\lambda},~\Omega_{\nu}) = (1,~0,~0)$, (0.4, 0.6,
0), and (0.7, 0, 0.3). ($\Omega_M$, $\Omega_{\lambda}$ and
$\Omega_{\nu}$ represent the density parameters for matter, vacuum
energy and the hot Dark Matter component, respectively.)  The maximum
number of lenses in a single plane is approximately 600, each having
the appropriate computed mass value, and they follow $1.85 \times
10^7$ uniformly distributed rays within a solid angle of $2.8 \times
10^{-6}$sr, corresponding to a $420'' \times 420''$ field. The final
impact parameters of the rays are collected in an orthogonal grid of
$300^2$ pixels in the source plane. Because of the use of point
masses, their method produces very high magnification values, greater
than 30 for the Einstein-de Sitter cosmology. They have also chosen to
use a smoothness parameter $\bar{\alpha} =0$ in the redshift-angular
diameter distance relation (described in Section 4) which depicts an
entirely clumpy universe.

Jain, Seljak and White (1998, 1999) have made use of $N$-body
simulations generated using a parallel adaptive particle-particle,
particle-mesh (AP$^3$M) code. The cosmologies simulated are the same
as the ones reported on here, but they have 256$^3$ particles and
comoving box side dimensions of 85$h^{-1}$Mpc for their Einstein-de
Sitter cosmologies, and 141$h^{-1}$Mpc for the open and cosmological
constant models. For each plane, the projected density, together with
the appropriate redshift-dependent factors, is Fourier transformed,
using periodic boundary conditions, to obtain the shear in Fourier
space. The two-dimensional shear matrix is then computed onto a grid
in real space by the inverse Fourier transform.  The size of the grid
is chosen to be compatible with the force-softening scale (the
resolution) in the $N$-body simulations, so that for sources at
redshifts around 1, the angular-scale size of the grid is less than,
or of the same order as, the angular-scale size of the force-softening
at these redshifts. Perturbations on the photon trajectories are
computed and the shear matrix interpolated to the photon positions
enabling the Jacobian matrices to be computed by recursion. In view of
the use of a fine grid for the shear and deflection angle
computations, they are able to analyse their data on different scales,
and are thus able to determine the power spectrum in both the shear
and the convergence. This approach is very different from the approach
we follow which makes use of variable softening for the particles
depending on their particular environment, and which ensures that most
`rays' would pass entirely through regions of smoothed density,
thereby requiring the application of the full beam
approximation. Jain, Seljak and White (1998, 1999) assume point
particles interpolated onto their grid, and use the empty cone
approximation.

Hamana, Martel and Futamase (1999) study weak lensing in $N$-body
simulations of three cosmologies with ($\Omega_0,~\lambda_0$) = (1,
0), (0.3, 0.7), and (0.3, 0). (We now use $\Omega_0$ and $\lambda_0$
to represent the present matter density and vacuum energy density
parameters respectively.) These were generated by a P$^3$M algorithm,
using Fourier techniques on a 128$^3$ lattice. The comoving simulation
box sizes were 128Mpc, and the particles were given comoving
softenings of 300kpc. Three simulations were performed for each
cosmological model, and boxes from the different simulations combined
to limit correlations in large-scale structure between adjacent
boxes. Each of the mass distributions in the simulation boxes were
than projected onto planes at the box redshift, and Poisson's equation
solved numerically on each of them. This was done by first evaluating
the surface density onto a $512~\times~512$ grid, based on the
particle positions, and then inverting Poisson's equation using a Fast
Fourier Transform (FFT) method. As the multiple lens-plane theory was
to be used for more than $10^7$ rays passing through each plane, the
recursion algorithm was simplified by assuming small deflections for
the rays. This also meant that the rays could be considered to pass
through the grid points, and travel, effectively, in straight lines
through the entire distance from the observer to the source plane. We
have also made this approximation in our own method.

An alternative to the conventional form of `ray-tracing' was
introduced by Refsdal (1970), who used `ray-tracing' with calculations
of the differential deflections of light rays around a central ray to
determine the distribution of magnifications.

Fluke, Webster and Mortlock (1999) and Fluke, Webster and Mortlock (2000) have
developed this idea further as the `ray-bundle' method. The principle
is to trace the passage of a discrete bundle of light rays as it
passes through the deflection planes. The advantage of the method is
that it provides a direct comparison between the shape and size of the
bundle at the observer and at the source plane, so that the
magnification, ellipticity and rotation can be determined
straightforwardly. The authors have selected a range of popular
cosmologies, including the ones reported on here, and have produced
their own $N$-body simulation data-sets using a P$^3$M algorithm. For
each cosmology they have produced a number of independent simulations
to enable them to randomly choose a simulation box for a particular
epoch from any of the realisations. In this way, correlations of
large-scale structure between adjacent boxes are avoided. They have
run their simulations with $64^3$ dark matter particles, with box
sizes ranging from 80$h^{-1}$Mpc to 164.3$h^{-1}$Mpc. Having randomly
selected the boxes for a given cosmology, the particle mass
distributions are then projected onto planes at the redshifts of the
boxes. The authors construct approximately 50,000 bundles, each
comprising 8 rays, to shoot in random directions through the planes,
from the observer's location at $z = 0$. The shooting area was limited
to $50'' \times 50''$ to avoid edge effects of the planes, and only
matter within a single period in the transverse direction is included
in the planes. In addition, a radius, typically 15$h^{-1}$Mpc centred
around each bundle is chosen for the extent of the matter to be
included in the deflection angle computations. The projected masses
are also considered as point particles, so that very high
magnification values can be achieved in principle; however, the
authors do not include bundles which pass within $\sqrt{2}$ of the
Einstein radius of any particle. Because of their use of point masses,
they use the empty cone approximation in the determination of the
statistical distributions of magnifications for the different
cosmological models.

Premadi, Martel and Matzner (1998a) have introduced individual
galaxies into the computational volume, matching the 2-point
correlation function for galaxies. They also assign morphological
types to the galaxies according to the individual environment, and
apply a particular surface density profile for each. Five different
sets of initial conditions were used for the simulations, so that the
individual plane projections can be selected at random from any
set. $N$-body simulations were produced for three cosmologies with
($\Omega_0,\lambda_0$) = (1, 0), (0.2, 0), and (0.2, 0.8) in boxes of
comoving side-dimension 128$h^{-1}$Mpc. They solve the two-dimensional
Poisson equation on a grid and invert the equation using a FFT method
to obtain the first and second derivatives of the gravitational
potential on each plane. They also correctly ensure that the mean
surface density in each lens-plane vanishes, so that a good
interpretation of the effects of the background matter is made. Their
method uses beams of light, each comprising 65 rays arranged in two
concentric rings of 32 rays each, plus a central ray. The multiple
lens-plane theory then enables the distributions of cumulative
magnifications to be obtained.

Tomita (1998a and b) also uses a ray bundle method, but by evaluating
the gravitational potential at approximately 3000 locations between
the observer and sources at redshift 5, he is able to compute the weak
lensing statistics without using the multiple lens-plane theory. He
has used $N$-body simulations produced using a tree code with $32^3$
particles in four different cosmologies with $(\Omega_0,~\lambda_0) =
(1,~0),~(0.2,~0.8),~(0.2,~0),$ and (0.4, 0). In all of these, except
the SCDM cosmology ($\Omega_0 = 1,~\lambda_0 = 0$), the particles are
treated as galaxy-size objects. In the SCDM cosmology, 20\% of the
particles are treated thus, whilst 80\% are given softening radii of
100$h^{-1}$kpc (model a), 40$h^{-1}$kpc (model b), and 20$h^{-1}$kpc
(model c). Ray bundles are formed from $5 \times 5$ rays arranged in a
square, with separation angles (fixed for each run) at between 2
arcsec and 1 degree. At each of the 3000 positions between the
observer and source, the potential is evaluated at each ray position
by translating each simulation cube (using the periodic properties) so
that the centre of each bundle is located in the centre of each
cube. In this way, all the matter within a full period in extent
contributes to the calculation of the potential (although no account
is taken of matter beyond one period transverse to the line of
sight). To avoid spurious values of the potential arising from masses
close to evaluation positions, an average of the potential is taken by
integrating it analytically over the interval between adjacent
evaluation positions. The light propagation is then determined by
solving the null-geodesic equations, and the required statistics
constructed from shooting 1000 bundles through the flat cosmologies,
and 200 bundles through the open cosmologies.

\section{THE PROPAGATION OF LIGHT}

In the case of multiple deflections of light by a series of projected
lens-planes, the Jacobian matrix develops in accordance with the
multiple lens-plane theory, which has been summarised by Schneider, Ehlers and
Falco (1992). Its form at redshift $z = 0$ enables all the lensed
properties of light from distant sources to be determined.

We follow, in outline, their description. For a system of $N$ lenses,
the basic lens equation for a single lens, relating the angular
position of the source to that of the image, may be generalised
straightforwardly. If the position vector of the source in the plane
perpendicular to the line of sight at the source position (the source
plane) is $\mbox{\boldmath$\eta$}$, and the position vectors of the
image positions in the various, $N$, deflection planes are
$\mbox{\boldmath$\xi_i$}$, where $i = 1,...,N$, then the lens equation
may be written as
\begin{equation}
\mbox{\boldmath$\eta$} = \frac{D_s}{D_1}\mbox{\boldmath$\xi_1$}-\sum_{i=1}^N D_{is}\hat{\mbox{\boldmath$\alpha$}}_i(\mbox{\boldmath$\xi_i$}),
\label{eta}
\end{equation}
where $D_i$ is the angular diameter distance to the $i$th lens,
$D_{is}$ is that from the $i$th lens to the source, and
$\hat{\mbox{\boldmath$\alpha$}}_i$ is the deflection angle at the
$i$th lens. To make equation~\ref{eta} dimensionless, put
\begin{equation}
\vec{x}_i = \mbox{\boldmath$\xi$}_i/D_i,
\label{xi}
\end{equation}
and for the source,
\begin{equation}
\vec{y} = \vec{x}_{N+1} = \mbox{\boldmath$\eta$}/D_s = \mbox{\boldmath$\eta$}/D_{N+1}.
\label{y}
\end{equation}
We also use the individual reduced deflection angles defined by
\begin{equation}
\mbox{\boldmath$\alpha$}_i = \frac{D_{is}}{D_s}\hat{\mbox{\boldmath$\alpha$}}_i.
\label{alphai}
\end{equation}
Then the displacement in the $j$th lens plane is
\begin{equation}
\vec{x}_j = \vec{x}_1-\sum_{i=1}^{j-1}\frac{D_{ij}}{D_j}\hat{\mbox{\boldmath$\alpha$}}_i(D_i\vec{x}_i) = \vec{x}_1-\sum_{i=1}^{j-1}\frac{D_s}{D_{is}}\frac{D_{ij}}{D_j}\mbox{\boldmath$\alpha$}_i(D_i\vec{x}_i), 
\label{xj1}
\end{equation}
or
\begin{equation}
\vec{x}_j = \vec{x}_1-\sum_{i=1}^{j-1}\beta_{ij}\mbox{\boldmath$\alpha$}_i(D_i\vec{x}_i), 
\label{xj2}
\end{equation}
where
\begin{equation}
\beta_{ij} \equiv \frac{D_s}{D_{is}}\frac{D_{ij}}{D_j}.
\label{beta}
\end{equation}
Then the full form of the ray-tracing equation is equation~\ref{xj2} evaluated at the source plane:
\begin{equation}
\vec{y} = \vec{x}_1-\sum_{i=1}^N\mbox{\boldmath$\alpha$}_i.
\label{raytracing}
\end{equation}
(The $\beta_{ij}$ factor disappears, because $\beta_{is} = 1$ from its
definition.) The mapping of the source onto the image is given by the
Jacobian matrix, which relates small changes in the source to
corresponding small changes in the image seen on the first lens-plane:
\begin{equation}
{\cal A} \equiv \frac{\partial \vec{y}}{\partial \vec{x} _{1}}.
\label{Jacobian4}
\end{equation}
Similarly, we may define Jacobian matrices appropriately at each of the
lens-planes:
\begin{equation}
{\cal A} _{i} \equiv \frac{\partial \vec{x} _{i}}{\partial \vec{x} _{1}}.
\label{Jacobian5}
\end{equation}
We now define the derivative of the reduced deflection angle for the
$i$th lens by
\begin{equation}
{\cal U} _{i} \equiv \frac{\partial \mbox{\boldmath$\alpha$} _{i}}{\partial \vec{x} _{i}}.
\label{U1}
\end{equation}
By defining the `effective lensing potential' in terms of the angular position, 
$\mbox{\boldmath$\theta$}_i$ for the $i$th lens as an integral along the line of sight direction, $x_3$,
\begin{equation}
\psi_i(\mbox{\boldmath$\theta$}_i) = \frac{D_{ds}}{D_dD_s}\frac{2}{c^2}\int\phi(D_d\mbox{\boldmath$\theta$}_i,x_3)dx_3,
\label{psi1}
\end{equation}
in which $c$ is the velocity of light, ${\cal U}_{i}$ can be shown to
be equivalent to the matrix of the second derivatives of the effective
lensing potential for the $i$th lens, and is therefore related to the
second derivative of the gravitational potential (the shear):
\begin{equation}
{\cal U} _{i} = \left( \begin{array}{cc}
	\psi_{11}^i  & \psi_{12}^i \\
	\psi_{21}^i   & \psi_{22}^i
	\end{array}
	\right),
\label{U2}
\end{equation}
where the superscripts, $i$, denote the deflection plane index, and
where we have written
\begin{equation}
\psi_{11} \equiv \frac{\partial^2\psi(\mbox{\boldmath$\theta$})}{\partial\theta_1^2},~\psi_{12} \equiv \frac{\partial^2\psi(\mbox{\boldmath$\theta$})}{\partial\theta_1\partial\theta_2}, \nonumber
\end{equation}
\begin{equation}
\psi_{21} \equiv \frac{\partial^2\psi(\mbox{\boldmath$\theta$})}{\partial\theta_2\partial\theta_1},~\mathrm{and}~\psi_{22} \equiv \frac{\partial^2\psi(\mbox{\boldmath$\theta$})}{\partial\theta_2^2},
\label{psixx}
\end{equation}
in which the suffixes in the denominators refer to the coordinate
directions. Then the ray-tracing equation (\ref{raytracing}) gives
\begin{eqnarray}
{\cal A} _{\mathrm {total}} \equiv \frac{\partial \vec{y}}{\partial \vec{x}_{1}} & = & {\cal I} -\sum _{i=1}^{N} \frac{\partial \mbox{\boldmath$\alpha$} _{i}}{\partial \vec{x} _{1}} \nonumber\\
& = & {\cal I} -\sum _{i=1}^{N} \frac{\partial \mbox{\boldmath$\alpha$} _{i}}{\partial \vec{x} _{i}}\frac{\partial \vec{x} _{i}}{\partial \vec{x} _{1}}={\cal I} -\sum_{i=1}^{N} {\cal U} _{i}{ \cal A} _{i},
\label{Jacobian6}
\end{eqnarray}
where $\cal I$ is the identity matrix. 

Thus the final Jacobian matrix
can be evaluated at $z=0$, since the individual matrices can be
obtained by recursion. Using equation~\ref{xj2}, they are just
\begin{equation}
{\cal A} _{j}={\cal I} -\sum _{i=1}^{j-1}\beta _{ij}{\cal U} _{i} {\cal A} _{i}
\label{Jacobian7}
\end{equation}
for the $j$th lens, and
\begin{equation}
{\cal A} _{1} = {\cal I}.
\label{A1}
\end{equation}

In our approach, the second derivatives of the two-dimensional
effective lensing potentials required for each deflection location,
are obtained by integration of the computed three-dimensional shear
values, i.e., the second derivatives of the {\em peculiar} gravitational
potential. It is necessary to work with the peculiar gravitational
potential, because the shearing of light arises from deviations from
homogeneity; in a pure Robertson-Walker universe we would expect no
deviations. Couchman, Barber and Thomas (1999) derive the expression
for the peculiar gravitational potential, $\phi$, in terms of the
gravitational potential, $\Phi$, and the mean density, $\bar{\rho}$:
\begin{equation}
\phi=\Phi - \fracnum{2}{3}\pi G a^2 \bar\rho x^2;
\label{8}
\end{equation}
$G$ is the universal gravitational constant, $\vec{x}$ is the position
vector, and $a$ is the expansion factor for the universe (so that
$a\vec{x}$ is the comoving position vector).  This result corresponds
to a system with zero net mass on large scales and immediately gives
\begin{equation}
{\partial\phi\over \partial x_i} = {\partial\Phi\over \partial x_i} -
\fracnum{4}{3}\pi G a^2 \bar\rho x_i
\label{9}
\end{equation}
and
\begin{equation}
{\partial^2\phi\over \partial x_i\partial x_j} = {\partial^2\Phi\over
\partial x_i\partial x_j}  - \fracnum{4}{3}\pi G a^2
\bar\rho \,\delta_{ij}.
\label{10} 
\end{equation}
Equation~\ref{10} shows how the real result for the shear,
${\partial^2\phi/ \partial x_i\partial x_j}$, based on the peculiar
gravitational potential, is related to the value of ${\partial^2\Phi/
\partial x_i\partial x_j}$ through the subtraction of the term in the mean density.

Equation~\ref{10} can now be evaluated explicitly for the
three-dimensional shear. The integral solution of Poisson's equation
is well-known, and the solution can be easily differentiated twice to
give
\begin{eqnarray}
\lefteqn{\frac{\partial^2\phi(\vec{R})}{\partial x_i\partial x_j} =} \nonumber\\
\lefteqn{G\int\!\!\!\int\!\!\!\int\left[\frac{\rho(\vec{R}')}{\mid\vec{R}-\vec{R}'\mid^3}\delta_{ij} 
-\frac{3\rho(\vec{R}')(x_i-x_i')(x_j-x_j')}{\mid\vec{R}-\vec{R}'\mid^5}\right]d^3R'} \nonumber\\
 & & \hskip 1.5 in -\fracnum{4}{3}\pi G a^2 \bar\rho \,\delta_{ij}.
\label{3dshear}
\end{eqnarray}
(We have introduced $\vec{R}$ and $\vec{R}'$ for the evaluation
position for the shear, and the matter positions respectively. In
practice, of course, the triple integral over all space would be
replaced by a summation.) The two-dimensional second derivatives of
the effective lensing potentials required for the Jacobian matrices
then follow immediately from equation~\ref{psi1} (using spatial rather
than angular coordinates):
\begin{equation}
\psi_{ij} = \frac{D_d D_{ds}}{D_s}.\frac{2}{c^2} \int\frac{\partial^2
\phi(x_3)}{\partial x_i \partial x_j}dx_3.
\label{psiij}
\end{equation}
$D_d$, $D_{ds}$, and $D_s$ are the angular diameter distances from the
observer to the lens, the lens to the source, and the observer to the
source, respectively. The integration is along the coordinate
direction, $x_3$, and the subscripts $i$ and $j$ now refer to any of
the three coordinate directions. This equation applies quite generally
for any deflection location, so the deflection plane index has been
dropped for clarity.

We now summarise the main equations we shall need for weak lensing,
which are obtainable from the final Jacobian matrix. The final
emergent magnification, $\mu$, may be computed after passage through
an entire box or set of boxes, and is
\begin{equation}
\mu =\left(\det \cal A \right)^{-1}.
\label{mu1}
\end{equation}

The convergence, $\kappa$, is
\begin{equation}
\kappa = \frac{1}{2}(\psi_{11}+\psi_{22}),
\label{kappa1}
\end{equation}
and is therefore obtainable from the diagonal elements of the Jacobian
matrix. 

The two-dimensional shear, $\gamma$, in each line of sight, is given by
\begin{equation}
\gamma^2 \equiv
\frac{1}{4}(\psi_{11}-\psi_{22})^2 + \psi_{12}^2.
\label{gammadef}
\end{equation}
(We may take $\psi_{12} = \psi_{21}$, because we are dealing with
a weak shear field which is smoothed by the variable particle
softening, ensuring that the gravitational potential and its
derivatives are well-behaved continuous functions.) 

From equation~\ref{mu1}, and these definitions, 
\begin{equation}
\mu = (1-\psi_{11}-\psi_{22}+\psi_{11}\psi_{22}-\psi_{12}^2)^{-1},
\end{equation}
or
\begin{equation}
\mu = \frac{1}{(1-\kappa)^2-\gamma^2}.
\label{musim}
\end{equation}
In the presence of convergence and shear, a circular source becomes
elliptical in shape, and the ellipticity, $\epsilon$, defined in terms
of the ratio of the minor and major axes, becomes
\begin{equation}
\epsilon = 1 - \frac{1-\kappa -\gamma}{1-\kappa +\gamma},
\end{equation}
which reduces to
\begin{equation}
\epsilon \simeq 2 \gamma (1+\kappa -\gamma ) + O(\kappa^3,\gamma^3)
\label{epssim}
\end{equation}
in weak lensing.

In our method, the evaluation of the second derivatives of the
two-dimensional effective lensing potentials is obtained from
integration of the computed three-dimensional shear values at a large
number of evaluation positions along lines of sight. The multiple
lens-plane theory then enables distributions of the magnification,
ellipticity, convergence and shear at redshift $z=0$ to be computed for light
rays traversing the set of linked simulation boxes starting from the
chosen source redshift. The ability to apply the appropriate angular
diameter distances at every evaluation position avoids the
introduction of errors associated with planar methods, and also allows
the possibility of choosing source positions within a simulation box
if necessary.

\section{PROCEDURE}
\subsection{The three-dimensional shear algorithm}

Couchman et al. (1998) describe in detail the algorithm for the
computation of the elements of the matrix of second derivatives of the
gravitational potential. The algorithm is based on the standard P$^3$M
method (see Hockney and Eastwood, 1988), and uses a Fast Fourier
Transform convolution method. It computes all of the six independent
shear component values at each of a large number of selected
evaluation positions within a three-dimensional $N$-body particle
simulation box. It has a computational cost of order $N{\rm log}_2 N$,
where $N$ is the number of particles in the simulation volume, and for
ensembles of particles, used in typical $N$-body simulations, the rms
errors in the computed shear component values are typically $\sim
0.3\%.$ In addition, the shear algorithm has the following features.

{\bf A)}~ The algorithm uses variable softening designed to distribute
the mass of each particle within a radial profile depending on its
specific environment. By virtue of this facility, we have been able to
choose the softening such that light rays feel the existence of a
smooth mass distribution.  Each particle may be assigned its own
softening-scale parameter, depending on the particle number-density in
its environment. In this way, it can be used to minimise the effects
of isolated single particles, whilst the smoothed denser regions are
able to represent the form of the large-scale structure. The parameter
we have chosen to delineate the softening scale for each particle is
proportional to $l,$ where $2l$ is the radial distance to the
particle's 32nd nearest neighbour. The value of $l$ has been evaluated
for every particle by using a smoothed particle hydrodynamics (SPH)
density program for each simulation box, and is read in by the shear
code along with the particle position coordinates.

The maximum softening is allowed to be of the order of the mesh
dimension for isolated particles, which is defined by the regular grid
laid down to decompose the short- and long-range force
calculations. In this way individual isolated particles are unable to
strongly influence the computed shear values, in accordance with our
need to study the broad properties of the large-scale structure,
rather than the effects of individual particles, which are not
representative of physical objects.

To determine realistic values for the minimum softening scale, we
wanted to keep the incidence of strong lensing to a minimum, whilst at
the same time allowing a physical interpretation. Using the
dimensionless angular diameter distances in terms of the present value
of the Hubble parameter, $H_0$, i.e., $r_d$ ($\equiv
\frac{H_0}{c}D_d$), $r_{ds}$ ($\equiv \frac{H_0}{c}D_{ds}$), and $r_s$
($\equiv \frac{H_0}{c}D_{s}$), for the observer to the lens, the lens
to the source, and the observer to the source, respectively, the
Einstein radius, $R_E$, becomes
\begin{equation}
R_E = 8.6 \times 10^{-3}h^{-1}N^{\frac{1}{2}}
\left(\frac{r_dr_{ds}}{r_s}\right)^{\frac{1}{2}}~\mathrm{Mpc}
\label{Einstein2}
\end{equation}
for a cluster of $N$ particles, each of mass $1.29 \times 10^{11}$
solar masses (see Section 3.2), where $h$ is the Hubble parameter in
units of 100 km s$^{-1}$ Mpc$^{-1}$. Substituting values for the
angular diameter distance factors then gives a maximum value of $R_E =
2.84 \times 10^{-3}h^{-1}N^{\frac{1}{2}}$ Mpc for a source at redshift
$z_s = 1$ in the SCDM cosmology. This occurs for a lens at redshift
$z_d =0.29$. Thus, for a cluster of 1000 particles, $R_E =
0.089h^{-1}$ Mpc. For a source at redshift 3.6, the maximum value of
$R_E$ is $0.108h^{-1}$ Mpc and occurs for a lens at redshift
0.53. Thus by setting a working minimum value for the variable
softening of $0.1h^{-1}$ Mpc, we would rarely expect to see strong
lensing due to caustics in the simulations. In box units, the
softening is $10^{-3}\times (1+z)$, where $z$ is the box redshift.

The corresponding values for $R_E$ in the other cosmologies, for
$N=1000$, are as follows.

For $\Omega_0 = 0.3$, $\lambda_0 = 0$, the maximum value of $R_E$ is
$0.093h^{-1}$ Mpc for $z_s = 1$, and occurs for $z_d =0.32$. For $z_s
= 3.6$, the maximum value of $R_E$ is $0.115h^{-1}$ Mpc, and occurs
for $z_d =0.58$.

For $\Omega_0 = 0.3$, $\lambda_0 = 0.7$, the maximum value of $R_E$ is
$0.104h^{-1}$ Mpc for $z_s = 1$, and occurs for $z_d =0.36$. For $z_s
= 3.6$, the maximum value of $R_E$ is $0.142h^{-1}$ Mpc, and occurs
for $z_d =0.67$.

Consequently, we are able to ensure that the incidence of strong
lensing is kept to very low levels in all cosmologies, and that the
minimum softening is always greater than, or similar to, the maximum
value of the Einstein radius for a cluster of 1000 particles.  At the
same time, the softening scale is approximately of galactic
dimensions, giving a realistic interpretation to the choice. We have
also accounted for the effects of the different numbers of particles
per box in the different cosmological simulations. The variable
softening scale for each particle has been retained at the same level
in all the cosmologies, so that the same mass value is contained
within it. Thus, around small-particle clusters there would be
differences in the softening from cosmology to cosmology, but around
large clusters, where the shearing is likely to be most important, the
differences would be tiny.

Couchman et al. (1998) have investigated the sensitivity of weak lensing
results to the input softening, finding differences only in limited numbers of lines of sight at the high magnification end of the distributions.

{\bf B)}~ The shear algorithm works within three-dimensional
simulation volumes, rather than on planar projections of the particle
distributions, so that angular diameter distances to every evaluation
position can be applied. It has been shown (Couchman et al., 1998)
that in specific circumstances, the results of two-dimensional planar
approaches are equivalent to three-dimensional values integrated
throughout the depth of a simulation box, provided the angular
diameter distance is assumed constant throughout the depth. However,
by ignoring the variation in the angular diameter distances throughout
the box, errors up to a maximum of 9\% can be reached at a redshift of
$z=0.5$ for SCDM simulation cubes of comoving side
100$h^{-1}$Mpc. (Errors can be larger than this at high and low
redshift, but the angular diameter distance multiplying factor for the
shear values is greatest here for sources we have chosen at a redshift
of 4.)

{\bf C)}~ The shear algorithm automatically includes the contributions
of the periodic images of the fundamental volume, essentially creating
a realisation extending to infinity. Couchman et al. (1998) showed
that it is necessary to include the effects of matter well beyond the
fundamental volume in general (but depending on the particular
particle distribution), to achieve accurate values for the
shear. Methods which make use of only the matter within the
fundamental volume may suffer from inadequate convergence to the
limiting values.

{\bf D)}~ The method uses the peculiar gravitational potential,
$\phi$, through the subtraction of a term depending upon the mean
density. Such an approach is equivalent to requiring that the net
total mass in the system be set to zero, and ensures that we deal only
with light ray deflections arising from departures from homogeneity.

\subsection{The Hydra {\em N}-body simulations}

The three-dimensional shear code can be applied to any
three-dimensional distribution of point masses confined within a cubic
volume, and produces shear values as if the fundamental volume were
repeated indefinitely to represent a three-dimensional periodic
distribution of masses. Each particle may be assigned an individual
mass, although in the tests and our application of the code, all the
particles were assumed to be dark matter particles with the same mass.

The code has been applied to the data bank of cosmological $N$-body
simulations provided by the Hydra Consortium
(http://hydra.mcmaster.ca/hydra/index.html) and produced using the
`Hydra' $N$-body hydrodynamics code (Couchman, Thomas and Pearce,
1995). Simulations from four different cosmologies were used, which
will be referred to as the SCDM, TCDM, OCDM and LCDM cosmologies. Each
of the simulations used a Cold Dark Matter-like spectrum, and the
parameters used in the generation and specification of these
cosmological simulations are listed in Table~\ref{cosmo}. 
\begin{table}
\begin{center}
\begin{tabular}{|c|c|c|c|c|c|c|}
\hline
Cosmology & $\Omega_0$ & $\lambda_0$ & $\Gamma$ & $\sigma_8$ & No. of    & Box side \\
          &            &             &          &            & particles & ($h^{-1}$Mpc) \\ 
\hline 
SCDM & 1.0 & 0.0 & 0.50 & 0.64 & $128^3$ & 100 \\ 
\hline 
TCDM & 1.0 & 0.0 & 0.25 & 0.64 & $128^3$ & 100 \\ 
\hline 
OCDM & 0.3 & 0.0 & 0.25 & 1.06 & $86^3$ & 100 \\ 
\hline 
LCDM & 0.3 & 0.7 & 0.25 & 1.22 & $86^3$ & 100 \\
\hline
\end{tabular}
\end{center}
\caption{Parameters used in the generation of the four different cosmological simulations.}
\label{cosmo}
\end{table}

$\Omega_0$ and $\lambda_0$ are the present-day values of the density
parameter and the vacuum energy density parameter respectively, so
that the SCDM and TCDM cosmologies are representative of Einstein-de
Sitter universes, whereas the OCDM cosmology represents a low density,
open universe, and the LCDM a low density, but spatially-flat universe
with a cosmological constant. The power spectrum shape parameter,
$\Gamma$, is set to 0.5 in the SCDM cosmology, but the empirical
determination (Peacock and Dodds, 1994) of 0.25 for cluster scales has
been used in the other cosmologies. In each case, the normalisation,
$\sigma_8$, on scales of $8h^{-1}$Mpc has been set to reproduce
the number density of clusters (Viana and Liddle, 1996).

In the SCDM and TCDM cosmologies, the number of particles is $128^3$,
leading to individual dark matter particle masses of
$M_{\mathrm{part}} = 1.29 \times 10^{11}h^{-1}$ solar masses. In the
low density universes, the number of particles is 0.3 times the number
in the critical density universes, leading to the same individual
particle masses. The simulation output times were chosen so that
consecutive simulation boxes may be snugly abutted; the
side-dimensions are $100h^{-1}$Mpc in every case. Consequently, there
are different numbers of time-outputs to a given redshift value for
the different cosmologies. For a nominal source redshift of $z_s = 4$
(which is the furthest extent of our weak lensing analysis), 33
simulation boxes were abutted to a redshift of 3.90 in the SCDM
cosmology, 33 to a redshift of 3.93 in the TCDM cosmology, 41 to a
redshift of 4.00 in the OCDM cosmology, and 48 to a redshift of 3.57
in the LCDM cosmology. (The {\em number} of boxes has no bearing on
the weak lensing statistics.)

\subsection{Orientation and lines of sight}

Each time-output in a given simulation run is generated using the same
initial conditions, so that a particular structure (although evolving)
occurs at the same location in all the boxes, and is therefore
repeated with the periodicity of the box. To avoid such obvious and
unrealistic correlations, we have arbitrarily translated, rotated (by
multiples of $90^{\circ}$) and reflected each time-slice, about each
coordinate axis, before linking them together to form the continuous
depiction of the universe back to the source redshift.

To follow the behaviour of light rays from a distant source through
the simulation boxes, and obtain distributions of the properties at
$z=0$, we construct a regular rectangular grid of directions through
each box. Since there are likely to be only small deflections, and the
point of interest is the statistics of output values, each light ray
is considered to follow one of the lines defined by these directions
through the boxes. The evaluation positions are specified along each
of these lines of sight.

We have selected 1000 evaluation positions on each of $100 \times 100$
lines of sight, which is well matched to the minimum variable
softening, giving adequate sampling in the line of sight direction;
the method has been tested using up to a total of $4 \times 10^6$
lines of sight, and we have found that whilst this smooths the
distribution plots for the magnification and shear at the high
magnification end and gives rise to higher maximum values of the
magnification, the statistical widths of the plots are virtually
unchanged. Since we are dealing with weak lensing effects and are
interested only in the statistical distribution of values, these lines
of sight adequately represent the trajectories of light rays through
each simulation box. It is sufficient also to connect each `ray' with
the corresponding line of sight through subsequent boxes in order to
obtain the required statistics of weak lensing. This is justified
because of the random re-orientation of each box performed before the
shear algorithm is applied.

\subsection{Conversion factors and approximations in the method}

The second derivatives of the two-dimensional effective lensing
potentials are obtained from the three-dimensional second derivatives
of the peculiar gravitational potential by integration, in accordance
with equation~\ref{psiij}. The integration of the three-dimensional
shear values has been made in small steps (0.02 of the box depth)
along each line of sight, enabling the weak lensing properties to be
determined from the Jacobian matrices and recorded at 50 evenly-spaced
locations along each line of sight in every simulation box. To
evaluate the absolute second derivatives of the effective lensing
potentials, the appropriate scaling factor is introduced, which
applies to the simulation box dimensions. From equation~\ref{psiij},
the factor $B(1+z)^2r_d r_{ds}/r_s$ can be extracted, where
$B=(c/H_0)(2/c^2)GM_{\mathrm{part}}\times(\mathrm{comoving~box~depth})^{-2}$.
For the simulation boxes used, which have comoving dimensions of $100
h^{-1}$Mpc, $B=3.733 \times 10^{-9}$, and the $(1+z)^2$ factor occurs
to convert the comoving code units into physical units. (The step
length for the integrations in each box is left as a code parameter,
so that it may be varied at will, although in the analysis reported
here, the step length was invariably 0.02~$\times$~the box depth.)

The procedure to obtain the two-dimensional second derivatives of the
effective lensing potentials involves a number of
approximations. 

First, it is assumed that the angular diameter
distances required for equation~\ref{psiij} vary linearly throughout
each step length (0.02~$\times$~the box depth). They are, however,
evaluated exactly at the 50 step positions through each box.

Second, in our approach we assume that, although each simulation
box is generated as a single simulation output-time representation,
the angular diameter distances vary throughout the depth of each box,
as they would in the real universe.

Third, as mentioned in the previous section, we assume only weak
deflections for light rays, so that they are assumed to follow the
straight lines of sight represented by the grid points of the
evaluation positions. Since we are interested only in the statistics
of lensing within large-scale structure simulations, this is a
perfectly acceptable practice. 

Finally, we have made use of a weak lensing approximation for the
computation of the intermediate Jacobian matrices. The full form for
the final Jacobian matrix (equation~\ref{Jacobian6}) is computed by
recursion, by including the intermediate matrices given by
equations~\ref{Jacobian7} and~\ref{A1}. Expanding
equation~\ref{Jacobian6} fully involves second and higher order terms,
arising from the cross multiplication of the intermediate Jacobian
matrices, the distance factors represented by $\beta_{ij}$, and the
matrices ${\cal U} _{i}$. For weak lensing in which all the $\psi_{ij}$
are much less than unity, the full form for $\cal A _{\mathrm {total}}$
simplifies considerably to
\begin{equation}
{\cal A} _{\mathrm{total}} \simeq {\cal I} - \sum _{i=1}^{N}{\cal U} _{i}.
\label{Aapprox}
\end{equation}
This is the form used; strong lensing events will still be recorded as
such, but using this approximation, the component values in $\cal A
_{\mathrm{total}}$ will not be accurate in strong lensing
cases. However, the incidence of strong lensing events is likely to be
very small, because of our choice of the softening scale (see Section
4.3). This being the case, any strong lensing events will not
adversely affect the weak lensing statistics determined in our
analysis.

\section{ANGULAR DIAMETER DISTANCES IN INHOMOGENEOUS UNIVERSES}
\subsection{The Dyer-Roeder equation}

Our three-dimensional approach allows the use of the appropriate
angular diameter distances at every single evaluation position. This
is not possible in two-dimensional approaches, where it is assumed
that all the lensing mass in a box is projected onto a plane at a single
angular diameter distance.

Since the angular diameter distances depend very much on the
distribution of matter and the particular cosmology, it is therefore
necessary to have available appropriate values for the angular
diameter distances for the particular distribution of matter in the
simulation data-set being investigated.

By considering the universe to be populated by randomly distributed
matter inhomogeneities, but resembling the Robertson-Walker,
Friedmann-Lema\^{i}tre model on large scales, a second order
differential equation is obtained for the angular diameter distance,
$D$, in terms of the density parameter, $\Omega_0$, for the
universe, the vacuum energy density parameter, $\lambda_0$, and the
redshift, $z$, of the source.  Dyer and Roeder (1973) made assumptions
about the type of matter distribution to obtain a more practical
equation for $\lambda_0 = 0$ cosmologies. They assumed that a mass
fraction, $\bar{\alpha}$, (called the smoothness parameter), of matter
in the universe is smoothly distributed, and that the fraction
$(1-\bar{\alpha})$ is bound into clumps. Then the equation for the
angular diameter distance (with $\lambda_0 =0$) is:
\begin{eqnarray}
\lefteqn{\left(z+1 \right) \left[\Omega_0 z+1\right]
\frac{d^2D}{dz^2}+\left[\frac{7}{2}\Omega_0 z + \frac{\Omega_0}{2} + 3
\right]\frac{dD}{dz}} \nonumber \\
 & & \hskip 0.7 in + \left[\frac{3}{2}\bar{\alpha}\Omega_0 + 
\frac{\mid \sigma \mid^2}{(1+z)^5} \right] D = 0,
\label{D1}
\end{eqnarray}
in which $\sigma$ is the optical scalar for the shear, introduced by
the matter distribution around the beam. 

In order to apply equation~\ref{D1}, Dyer and Roeder (1973) considered
the following scenarios. They considered a universe in which all the
matter is bound into clumps, so that $\bar{\alpha}=0$, and in which
the light beam passes far away from the clumps. This is described as
light propagating through an `empty cone,' and gives rise to maximal
divergence of the beam. The opposite scenario has $\bar{\alpha}=1$,
i.e., an entirely smooth universe. Here the smooth matter distribution
is present within the beam, giving a `full cone,' or `filled beam'
approximation.

We have considered whether the shear along individual lines of sight
is able to significantly affect the chosen values for the angular
diameter distances.  Our work has been conducted using cosmological
simulations in which the distributions of matter are smooth, and we
show in Section 5.2 that the effects are negligible.  With $\sigma
\sim 0$, therefore, equation~\ref{D1} immediately reduces to the
well-known Dyer-Roeder equation,
\begin{eqnarray}
\lefteqn{\left(z+1 \right) \left[\Omega_0 z+1\right]
\frac{d^2D}{dz^2}+\left[\frac{7}{2}\Omega_0 z + \frac{\Omega_0}{2} + 3
\right]\frac{dD}{dz}} \nonumber\\
 & & \hskip 1.3 in + \frac{3}{2}\bar{\alpha}\Omega_0 D = 0,
\label{DR}
\end{eqnarray}
which can be solved analytically for $\Omega_0 = 1$, $\lambda_0 = 0$,
and arbitrary $\bar{\alpha}$.

\subsection{Generalisation of the Dyer-Roeder equation}

\begin{figure}
$$\vbox{
\psfig{figure=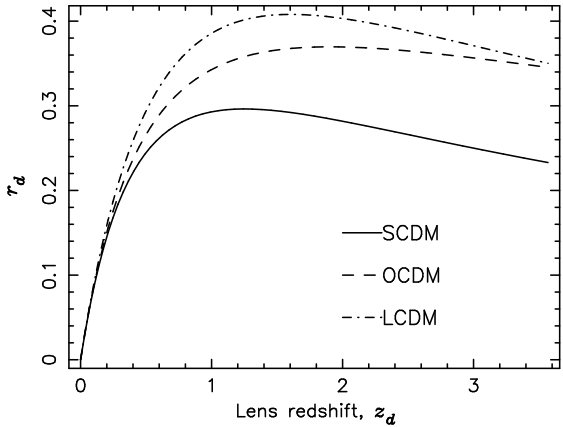,width=8.7truecm,angle=270}
}$$
\caption{The angular diameter distance, $r_d$ for the different cosmologies, assuming a source redshift of $z_s = 3.6$, and $\bar{\alpha} = 1$.} 
\label{dyeroeder1}
\end{figure}

\begin{figure}
$$\vbox{
\psfig{figure=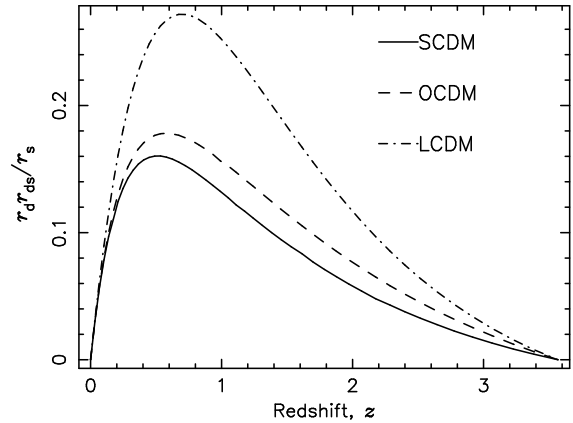,width=8.7truecm,angle=270}
}$$
\caption{The angular diameter distance multiplying factor, $r_dr_{ds}/r_s$ in the different cosmologies, assuming a source redshift of $z_s = 3.6$, and $\bar{\alpha} = 1$.} 
\label{dyeroeder2}
\end{figure}

Starting from the generalised beam equation, quoted by Linder (1998a
and b), we have generalised the form of the Dyer-Roeder equation to
apply to all the cosmologies we have simulated. This was necessary
because solutions for the angular diameter distances were required in
the LCDM cosmology containing a vacuum energy density. The procedure
to generalise the Dyer-Roeder equation is described fully in the
appendix, and solutions of the final equation were obtained
numerically.  Figure~\ref{dyeroeder1} shows the result of solving the
generalised equation, with $\bar{\alpha}=1$, in the different
cosmologies, for a source redshift of $z_s = 3.6$, and
Figure~\ref{dyeroeder2} shows the values of the angular diameter
distance multiplying factor, $r_dr_{ds}/r_s$ (which we now denote by
$R$), also for $\bar{\alpha}= 1$. It is clear from this plot that the
angular diameter distance multiplying factor is considerably higher in
the LCDM cosmology than the other cosmologies, and we shall comment
further on this in regard to the weak lensing statistics in the
discussion of our results in Section 7. When $\bar{\alpha}=0$, the
values are lower than for $\bar{\alpha}=1$; we have tabulated the
ratios $R(\bar{\alpha}=1)/R(\bar{\alpha}=0)$ for the different
cosmologies in Table~\ref{alphaR}.
%
%
%
%
%
%
%

\begin{table}
\begin{center}
\begin{tabular}{|l|c|c|}
\hline
Cosmology & Minimum $\bar{\alpha}$ &  $R(\bar{\alpha}=1)/R[\bar{\alpha}(z=0)]$ \\ 
\hline
SCDM & 0.83 & 1.0408 \\
TCDM & 0.88 & 1.0204 \\
OCDM & 0.80 & 1.0241 \\
LCDM & 0.82 & 1.0414 \\
\hline
\end{tabular}
\end{center}
\caption{The values of $\bar{\alpha}$ at $z = 0$ and the ratios of the $R$ factors at the peaks of the curves in the different cosmologies, assuming a source redshift of 3.6.}
\label{alphaR}
\end{table}

%
%
%
%
%
%
%
%
%

\subsection{Magnification in inhomogeneous universes}

Later, in Section 7, we make comparisons of our results with other
workers, who may use either the full beam or empty beam approaches for
the propagation of light. We find it difficult to make meaningful
comparisons with results obtained using the empty beam approach,
because the magnification distributions, for example, may be quite
different depending on the approach used. We therefore now consider
the effects of inhomogeneities, in the different approaches, which are
described loosely in terms of Dyer and Roeder's smoothness parameter,
$\bar{\alpha}$. We follow the line of reasoning given by Schneider,
Ehlers and Falco (1992).

Consider our inhomogeneous universe to be on-average, i.e., on large
scales, homogeneous and isotropic, so that the average flux from a
source at redshift $z$ and luminosity $L$ will equal the flux,
$S_{\mathrm{FL}}$, observed in a smooth Friedmann-Lema\^{i}tre
universe without local inhomogeneities:
\begin{equation}
\langle S \rangle = S_{\mathrm{FL}} = \frac{L}{4\pi[\bar{D}_L(z)]^2}.
\label{<S>}
\end{equation}
$\bar{D}_L(z)$ is the luminosity distance in the smooth
Friedmann-Lema\^{i}tre model, and is the mean of the luminosity
distance values in an inhomogeneous universe including the constraints
of flux conservation. Thus $\bar{D}_L(z)$ can be related to the
Dyer-Roeder angular diameter distance, $D_s$, in an entirely smooth
universe (with $\bar{\alpha} = 1$):
\begin{equation}
\bar{D}_L(z) = (1+z)^2D_s(z;\bar{\alpha}=1).
\label{D_L}
\end{equation}
Now the magnification, $\mu$, is just the ratio of the flux actually
observed in the image of a source and the flux which the same source
would produce if observed through an empty cone without
deflection. Then equation~\ref{<S>} straightaway gives, for the mean
magnification in terms of the appropriate Dyer-Roeder angular diameter distances,
\begin{equation}
\langle \mu \rangle = \left[\frac{D_s(z;\bar{\alpha})}{D_s(z;\bar{\alpha}=1)}\right]^2.
\label{<mu>}
\end{equation}
Clearly, the magnification values derived in this way depend on the
approximation used, and specifically the value of $\bar{\alpha}$. For
example, rays passing close to clumps or through high-density regions
will result in magnification in any approximation. If the empty cone
approximation is used, then $\mu$ will be greater than 1, and if the
full cone approximation is used, then $\mu$ will be greater than the
mean magnification. From equation~\ref{<mu>} it follows immediately
that the mean magnifications, $\langle \mu_e \rangle$ and $\langle
\mu_f \rangle$, in the empty cone and full beam approximations
respectively are
\begin{equation}
\langle \mu_e \rangle = \left[\frac{D_s(z;\bar{\alpha}=0)}{D_s(z;\bar{\alpha}=1)}\right]^2
\label{<mu_e>}
\end{equation}
and
\begin{equation}
\langle \mu_f \rangle = 1.
\label{<mu_f>}
\end{equation}
$\langle \mu_e \rangle \geq 1$ because $D_s(z;\bar{\alpha}=0) \geq
D_s(z;\bar{\alpha}=1)$ at all redshifts in all of our
cosmologies. This is an important result for advocates of the empty
beam approximation, particularly those working with distributions of
point mass particles, because by evaluating numerically the angular
diameter distance factors in the different cosmologies, (see the
Appendix), it immediately follows that
\begin{equation}
\langle \mu_e \rangle (\mathrm{OCDM}) \leq \langle \mu_e \rangle (\mathrm{LCDM}) \leq \langle \mu_e \rangle (\mathrm{SCDM}).
\label{comparison1}
\end{equation}
Pei (1993) succeeded in calculating the statistical properties of the
magnifications due to a random distribution of point mass lenses using
the assumption that the total magnification is a result of
multiplication of the magnifications produced at each redshift
interval. He found that the mean magnification as a function of
redshift was exponential in terms of the optical depth, $\tau(z)$:
\begin{equation}
\langle \mu_e(z) \rangle = \mathrm{exp}[2\tau(z)].
\label{Pei}
\end{equation}
(The optical depth is the fraction of the sky covered by circles of
Einstein radii between the observer and the specified redshift, and is
therefore dependent on the cosmology or the distribution and density
of matter. Pei (1993) gives an expression for the optical depth at
redshift $z$ in terms of the Dyer-Roeder smoothness parameter and the
cosmological density parameter for the point mass lenses in the SCDM
cosmology.) Although these are useful results, interpretation of the
magnification distributions in the empty cone approximation for
different cosmologies is often complicated by the high magnification
tails in the distributions which arise from light rays passing close
to point mass particles. Rauch (1991) set up a random distribution of
point masses and performed Monte Carlo simulations to calculate the
resulting amplification probability distributions. He then fitted the
distribution by an analytical expression for the probability which was
given only in terms of the mean magnification. Thus, for a given mean
magnification, the distribution curves would be identical, and almost
certainly somewhat unrealistic. However, since the mean magnification
for sources at a given redshift is dependent on the cosmology, the
actual distribution curves are able, in principle, to distinguish
between cosmologies.

The minimum magnification in the empty cone approximation is
\begin{equation}
\mu_{e,\mathrm{min}} = 1.
\label{mu_e,min}
\end{equation}
Thus rays passing through voids will have $\mu = \mu_{e,\mathrm{min}} =
1$ in the empty cone approximation, since the rays will be far from
all concentrations of matter, and will satisfy the empty cone
conditions. 

\begin{figure}
$$\vbox{
\psfig{figure=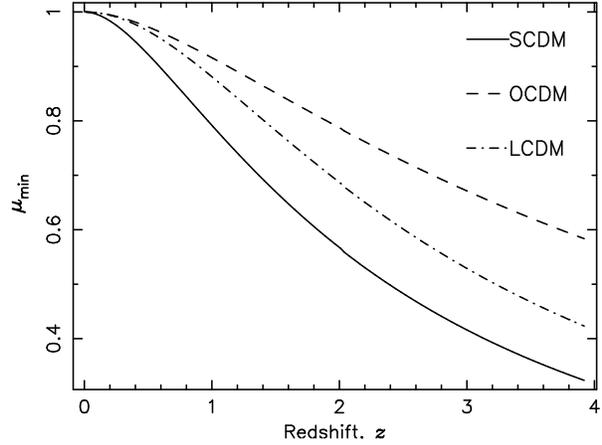,width=8.7truecm,angle=270}
}$$
\caption{The minimum magnification values versus redshift for 
the filled beam approximation in three different cosmologies.} 
\label{mumin}
\end{figure}

In the full cone approximation however, the magnification for rays
passing through voids will be less than or equal to 1 because the rays
will suffer divergence, and the minimum value will be
\begin{equation}
\mu_{f,\mathrm{min}} = \left[\frac{D(z;\bar{\alpha}=1)}{D(z;\bar{\alpha}=0)}\right]^2 \leq 1.
\label{mu_f,min}
\end{equation}
From this equation, we can again make comparisons amongst our
cosmologies. Figure~\ref{mumin} plots the value of
$\mu_{f,\mathrm{min}}$ from this equation versus redshift for the
different cosmologies, and immediately we can see that
\begin{equation}
\mu_{f,\mathrm{min}}(SCDM) \leq \mu_{f,\mathrm{min}}(LCDM) \leq \mu_{f,\mathrm{min}}(OCDM)
\label{comparison2}
\end{equation}
for all redshifts. 
%
%

There are two points to note about this result. First, the minimum
magnification values will only rarely (if ever) be seen, because this
would require the rays to pass entirely through the most underdense
regions. For this reason the values may only be treated as lower
bounds for the computed values. Comparisons between the computed
minima and these analytic values are shown in Table~\ref{Tmumin}. As
required, the computed values are consistently greater than the
theoretical minima. Second, the values say nothing about the
distribution of magnifications in any cosmology; it is not necessary,
for example, for the cosmology producing the lowest minimum
magnification to have the broadest range for the probability
distribution. Consequently, the results reported in Section 5 require
only that the minimum magnifications satisfy the minimum theoretical
bounds, and may not relate in any specific way to results of other
workers using point mass particles in an empty cone scenario. This
point is discussed further in Section 7.

\begin{table}
\begin{center}
\begin{tabular}{|ccc|}
\hline
Redshift & $\mu_{f,\mathrm{min}}$ (analytic) & $\mu_{f,\mathrm{min}}$ (computed) \\ \hline
\underline{SCDM/TCDM} & & \\ 
0.5 & 0.9245 & 0.9547 \\
1.0 & 0.7892 & 0.9030 \\
2.0 & 0.5656 & 0.8325 \\
4.0 & 0.3201 & 0.7486 \\ \hline
\underline{OCDM} & & \\ 
0.5 & 0.9734 & 0.9801 \\
1.0 & 0.9147 & 0.9510 \\
2.0 & 0.7862 & 0.8888 \\
4.0 & 0.4192 & 0.7922 \\ \hline
\underline{LCDM} & & \\ 
0.5 & 0.9662 & 0.9766 \\
1.0 & 0.9264 & 0.8791 \\
2.0 & 0.6854 & 0.8253 \\
4.0 & 0.5802 & 0.7145 \\ \hline
\hline
\end{tabular}
\end{center}
\caption{The analytical minimum magnification values in the filled 
beam approximation for the different cosmologies, and the 
computed values from the simulations. It is necessary only that 
the computed values should be larger than the minima, since the 
theoretical minima will not, in practice, be seen.}
\label{Tmumin}
\end{table}

\section{WEAK LENSING RESULTS IN THE DIFFERENT COSMOLOGIES}
\subsection{The formation of structure}

The formation of structure occurs at different rates in the different
cosmologies. Richstone, Loeb and Turner (1992), for example,
considered the spherical collapse of density perturbations, starting
from an initial Gaussian distribution, and found that the rate of
cluster formation as a function of redshift depended crucially on the
value of $\Omega_0.$ This has been confirmed by, for example,
Bartelmann, Ehlers and Schneider (1993). Lacey and Cole (1993),
starting from the basic Press-Schechter formul\ae, derived an equation
for the merger rates of virialised halos in hierarchical models, which
again showed the rates to be crucially dependent on $\Omega_0$. Later,
Lacey and Cole (1994) compared their analytical results with the
merger rates seen in $N$-body particle simulations for the SCDM
cosmology, and found good agreement for this cosmology. In addition,
their analytical result was applicable to arbitrary values of
$\Omega_0$, and more general power spectra.

Peebles (1993) summarises the evolution of structure in the
Press-Schechter approximation, which provides the number density for
collapsed objects by mass scale. This can be evaluated in terms of the
rms mass fluctuation, $\sigma_r$, at a fixed comoving scale, and so is
redshift dependent. For $\Omega_0$ close to 1, $\sigma_r \propto
(1+z)^{-1}$, and determines the evolution of the comoving number
density of clusters. On this model, half of the present-day number of
clusters would have formed later than $z \sim 0.1$, and 90\% would
have formed later than $z \sim 0.3$. This rapid evolution at late
times in an Einstein-de Sitter universe is seen also in $N$-body
simulations. In low density universes the time evolution of $\sigma_r$
is slower at low redshift, and this reduces the predicted rate of
cluster formation at late times. The Press-Shechter approximation also
underestimates the final number density for clusters. Carroll, Press
and Turner (1992) describe clearly the r\^{o}le of the cosmological
constant in the rate of structure formation. As $\Omega_0$ is reduced
from unity, the rate of growth is suppressed, but somewhat less so in
the presence of a cosmological constant. Thus, in the open case,
linear growth stops when $(1+z) \sim \Omega_0^{-1}$, when the universe
effectively becomes curvature dominated, but growth stops more
recently, when $(1+z) \sim \Omega_0^{-\frac{1}{3}}$, in the flat case,
when the universe effectively becomes dominated by the cosmological
constant.

There are clear qualitative differences between Einstein-de Sitter
universes and open models from $N$-body simulations. There is much
more dominance of clusters and groups of galaxies at earlier times in
the open models, which are then frozen in; however, open universes do
not display so prominently the large-scale filaments and other
irregular structures which occur in $\Omega_0 =1$ universes. Flat
cosmologies with a cosmological constant are intermediate between
these cases. It should be mentioned at this point that the results of
$N$-body simulations require normalisation against observations;
whilst the foregoing description of structure formation rates is not
very sensitive to the shape parameter, $\Gamma$, in the power
spectrum, both $\Gamma$ and the form of the initial conditions (for
example, Gaussian or non-Gaussian) may effect the resulting abundances
on different scales. For example, the SCDM model fails to reproduce
correctly the shape of the galaxy correlation function on scales of
tens of Megaparsecs, using $\Gamma = 0.5$, but the TCDM model, with
$\Gamma = 0.25$ does much better.

We now consider the values of the computed three-dimensional shear in
each time-slice, and their development with time, to see whether
comparisons may be made with the development of structure in the
different cosmologies. It is also informative to assess at what point
(or in which simulation boxes) the maximum contributions to properties
such as the magnification may occur. 

\begin{figure}
$$\vbox{
\psfig{figure=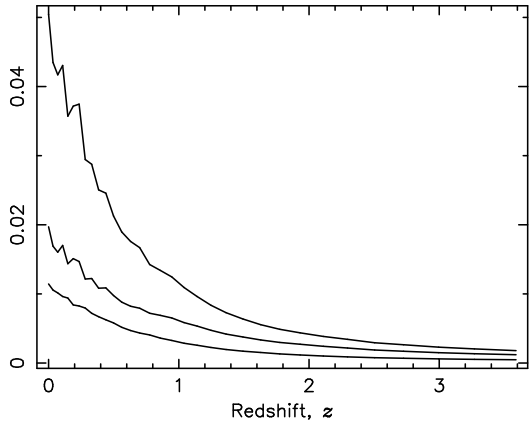,width=8.7truecm,angle=270}
}$$
\caption{Curves characterising the time-slices in the SCDM cosmology, 
derived directly from the computed values, and before applying any of 
the conversion factors. The ordinate is in arbitrary units.
Middle curve: the rms value in each time-slice of the sum of the first two
diagonal elements of the shear matrix. Top curve: the mean of the
high values of these elements. Lowest curve: the 
mean of the high values for one of the off-diagonal elements.}
\label{strS}
\end{figure}

\begin{figure}
$$\vbox{
\psfig{figure=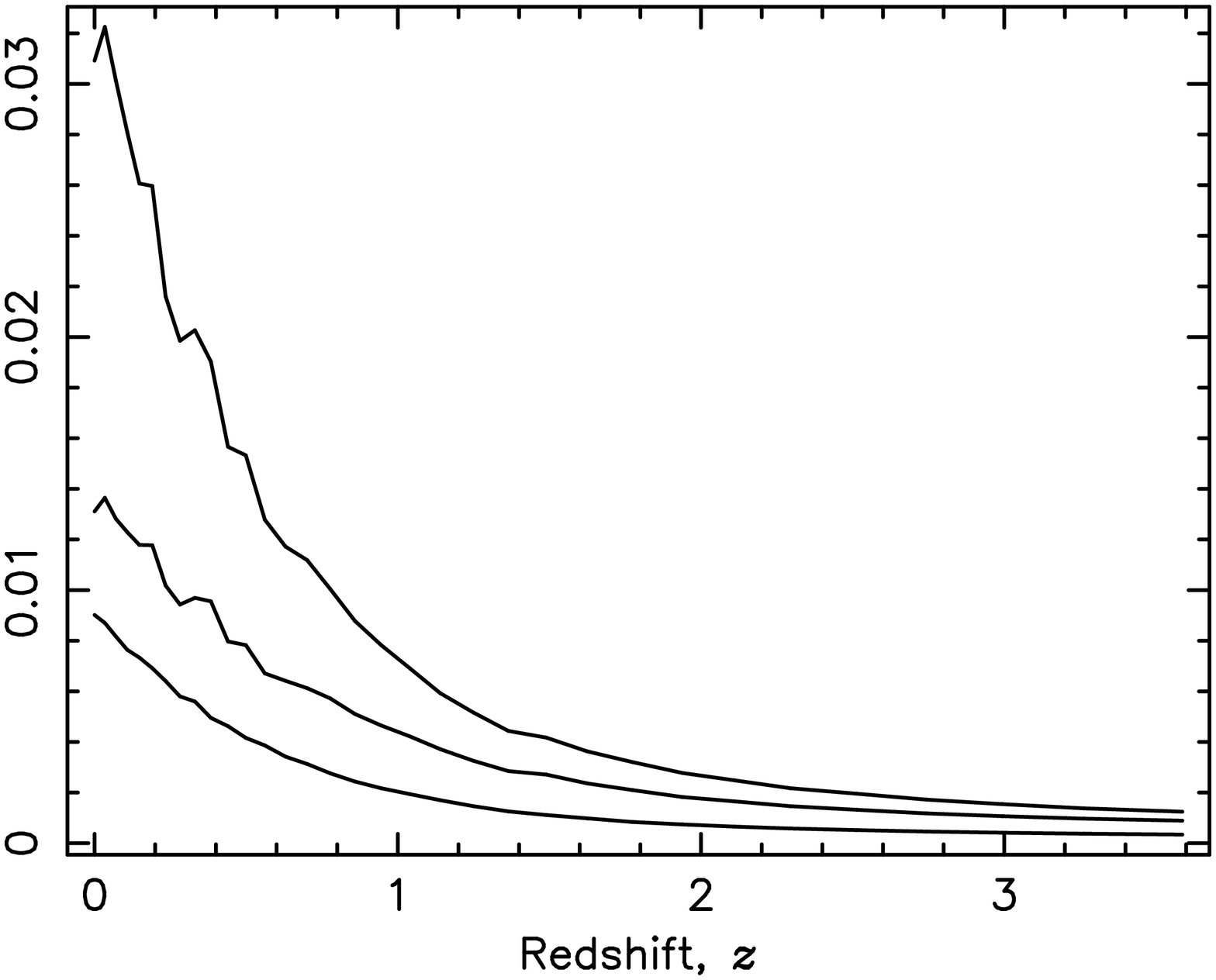,width=8.7truecm,angle=270}
}$$
\caption{Curves, as for Figure~\ref{strS}, characterising the time-slices
in the TCDM cosmology.} 
\label{strT}
\end{figure}

\begin{figure}
$$\vbox{
\psfig{figure=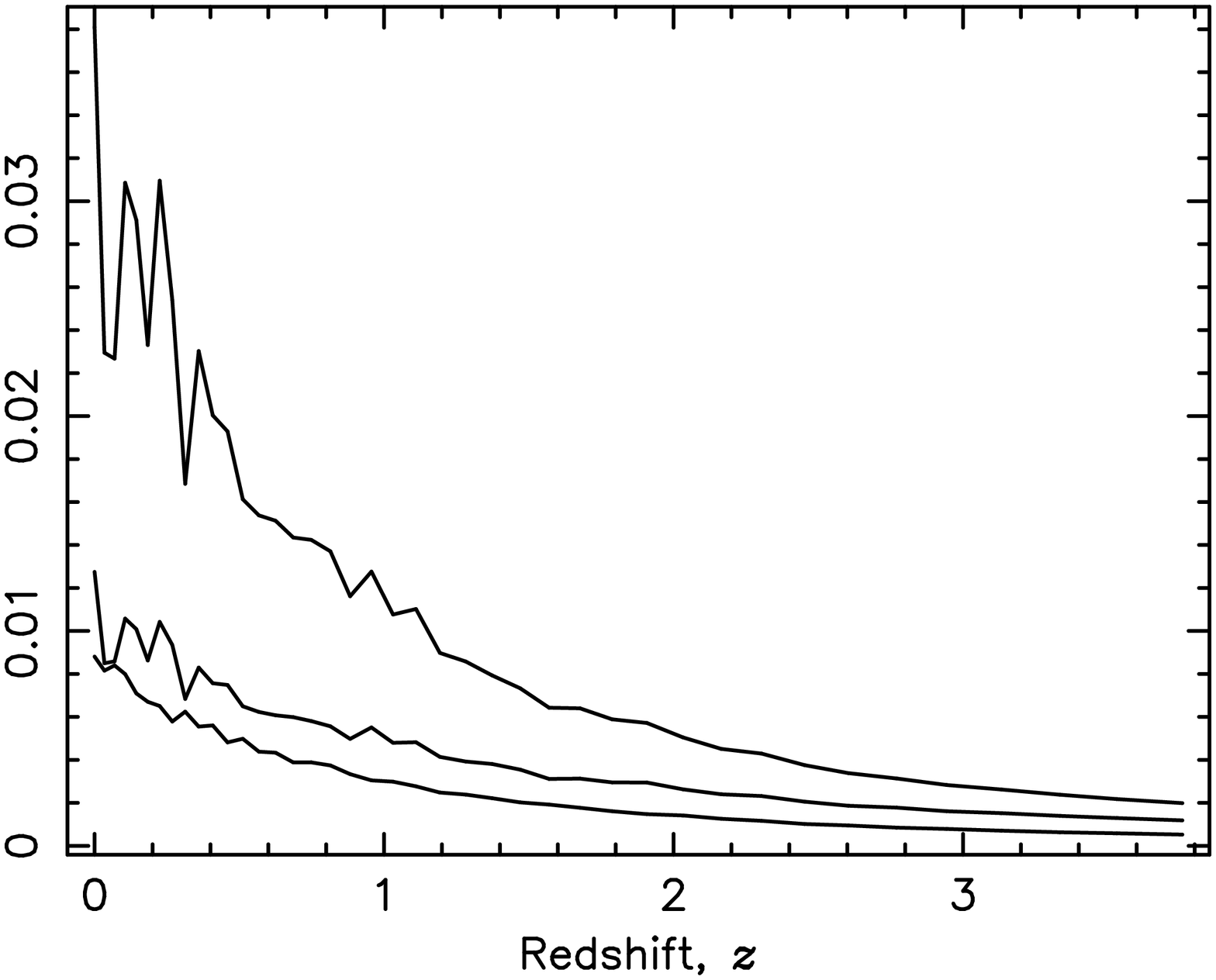,width=8.7truecm,angle=270}
}$$
\caption{Curves, as for Figure~\ref{strS}, characterising the time-slices
in the OCDM cosmology.} 
\label{strO}
\end{figure}

\begin{figure}
$$\vbox{
\psfig{figure=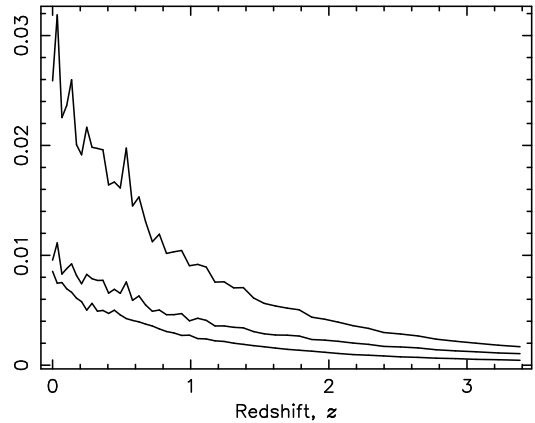,width=8.7truecm,angle=270}
}$$
\caption{Curves, as for Figure~\ref{strS}, characterising the time-slices
in the LCDM cosmology.} 
\label{strL}
\end{figure}

By simply taking the rms values of specified computed components, it
is possible, in a simplistic way, to obtain values which characterise
each time-slice, independent of its redshift. This is done (a) before
integration along the line of sight (which would be necessary to
obtain the two-dimensional effective lensing potentials, from which
the magnifications are obtained), (b) before conversion to physical
units (which involve factors to convert from the code units, and
factors which derive from the evolving box dimensions), and (c) before
the application of the appropriate angular diameter distance factors
(required in the application of the multiple lens-plane theory,
described in Section 2). Figures~\ref{strS},~\ref{strT},~\ref{strO}
and~\ref{strL} show the results for the SCDM, TCDM, OCDM and LCDM
cosmologies respectively. In each, the middle curve shows the rms
value in each time-slice of the sum of the first two diagonal elements
of the shear matrix, the top curve shows the mean of only the high
values of these elements, and the lowest curve shows the mean of the
high values for one of the off-diagonal elements. (The high values in
each case are those more than $1\sigma$ above the mean in the rms
values.) The growth in the values with time is clearly seen in each
cosmology.
%
%
%
%
%
%
%
%
%
%
%
%

\begin{figure}
$$\vbox{
\psfig{figure=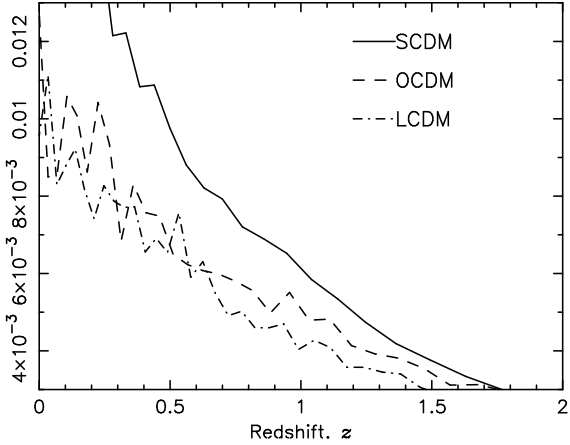,width=8.7truecm,angle=270}
}$$
\caption{Curves characterising the time-slices in the SCDM, OCDM and 
LCDM cosmologies, 
derived directly from the computed values, and before applying any of 
the conversion factors. The ordinate is in arbitrary units.
The curves derive from the rms values in each time-slice of the sum 
of the first two
diagonal elements of the shear matrix.}
\label{strComb1}
\end{figure}

\begin{figure}
$$\vbox{
\psfig{figure=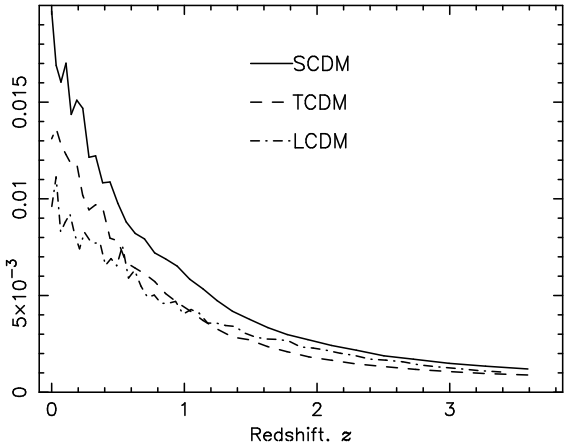,width=8.7truecm,angle=270}
}$$
\caption{Curves characterising the time-slices in the SCDM, TCDM and 
LCDM cosmologies, 
derived directly from the computed values, and before applying any of 
the conversion factors. The ordinate is in arbitrary units.
The curves derive from the rms values in each time-slice of the sum 
of the first two
diagonal elements of the shear matrix.}
\label{strComb2}
\end{figure}

In Figures~\ref{strComb1} and~\ref{strComb2} we have combined the data
for the different cosmologies. Figure~\ref{strComb1} shows the
directly computed rms values, as above, in the SCDM, OCDM and LCDM
cosmologies, and Figure~\ref{strComb2} the values in the SCDM, TCDM
and LCDM cosmologies. (There is some repetition here only to make the
comparisons clear, and to avoid too much overlapping of the curves.)
%
%
%
%
%
%
The plots in Figures~\ref{strComb1} and~\ref{strComb2} can be
understood in terms of the discussion at the beginning of this
section. The Einstein-de Sitter universes, SCDM and TCDM, show the
most rapid growth of the shear components at late times, reflecting
the rapid growth of structure; the OCDM and LCDM results would seem to
indicate only a limited effect from the cosmological constant.

\begin{figure}
$$\vbox{
\psfig{figure=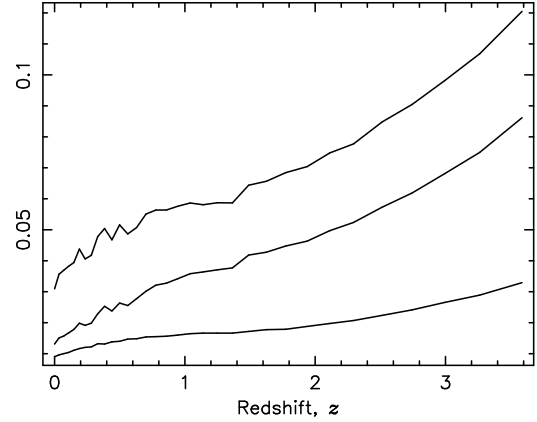,width=8.7truecm,angle=270}
}$$
\caption{Curves characterising the time-slices in the TCDM cosmology, 
obtained by multiplying the computed values by $(1+z)^3$ in each time-slice.
Middle curve: the rms value in each time-slice of the sum of the first two
diagonal elements of the shear matrix. Top curve: the mean of the
high values of these elements. Lowest curve: the 
mean of the high values for one of the off-diagonal elements.}
\label{strTA}
\end{figure}

Since the real physical dimensions of the simulation time-slices
evolve with time, it is necessary to introduce factors of $(1+z)^3$ to
the computed values to determine the real shear matrix. Doing this
appears to dilute considerably the effect of structure on the form of
the curves, as the example of Figure~\ref{strTA} for the TCDM
cosmology shows. (The curves are similar in form for all the
cosmologies.) The interpretation of this dilution is that, even though
structure is forming (to produce greater shear values locally), the
real expansion of the universe (causing the mean particle separation
to increase) reduces the shear values in most locations, and therefore
just outweighs the increases from the formation of structure. In this
way, the magnitudes of the shear component values are seen to reduce
slowly with time.
%
%
%

\begin{figure}
$$\vbox{
\psfig{figure=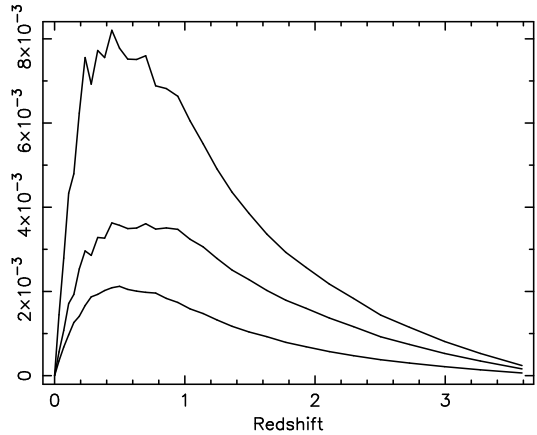,width=8.7truecm,angle=270}
}$$
\caption{The two-dimensional components in each time-slice (obtained 
by integration of the three-dimensional values), converted to absolute 
values,
including the angular diameter distance factors, for sources at
$z_s=4$. Middle curve: the rms value in each time-slice of the sum 
of the diagonal elements of the Jacobian matrix. Top curve: the mean 
of the high values of the summed diagonal elements. Lowest curve: 
the mean of the high values for one of the off-diagonal elements 
of the Jacobian.}
\label{str*RS}
\end{figure}

To obtain the two-dimensional effective lensing potentials (to which
the multiple lens-plane theory may be applied) it is necessary to
integrate the three-dimensional values, correctly converted to
physical units, along the line of sight, and to apply the appropriate
angular diameter distance factors in accordance with
equation~\ref{psiij}.  This requires the integrated three-dimensional
computed values (before any of the above factors are applied) to be
multiplied by the factor $B(1+z)^2r_dr_{ds}/r_s$, where
$B=(c/H_0)(2/c^2)GM_{\mathrm{part}}\times
(\mathrm{comoving~box~depth})^{-2}$, as described in Section 3.4.
When the computed values are multiplied by the full conversion factors
in this way, we see in Figure~\ref{str*RS} (for the SCDM cosmology,
and a source redshift of 3.9) that the peaks are extremely broad,
indicating that significant contributions to the magnifications and
ellipticities can arise in time-slices covering a wide range of
redshifts, and not just near $z=0.5$, where the angular diameter
distance multiplying factor, $R$, has its peak (for sources at
$z_s=4$). (In this exercise, we have used $\bar{\alpha} = 1$ for the
angular diameter distances. This is explained more fully in the next
section.)
%
%
%

\begin{figure}
$$\vbox{
\psfig{figure=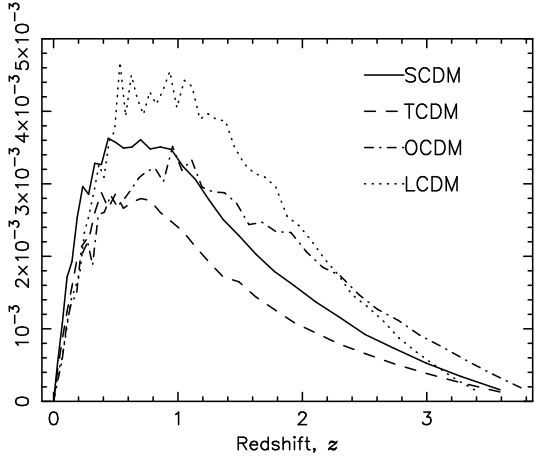,width=8.7truecm,angle=270}
}$$
\caption{The rms values of the sum of the diagonal elements of the 
Jacobian matrix in each time-slice (obtained by integration of the
three-dimensional values), converted to absolute values including the
angular diameter distance factors, for sources at $z_s=4$. Results are
shown for the SCDM, TCDM, OCDM, and LCDM cosmologies.}
\label{str*Rcomb}
\end{figure}

The comparisons amongst the cosmologies are
interesting. Figure~\ref{str*Rcomb} shows the integrated rms values
(as above) multiplied by the full conversion factors and the angular
diameter distance factors for the four cosmologies, assuming a source
redshift of 4. (The actual source redshifts differ very slightly in
the different cosmologies.) We see that the LCDM cosmology has both
the broadest and the highest peak, suggesting that lenses throughout
the broadest redshift range are able to contribute significantly to
magnifications and two-dimensional shearing of images in this
cosmology. Because of this, the magnitudes of the magnifications are
likely to be greatest in the LCDM cosmology. Significantly, the much
higher values for the angular diameter distance factor, $R$, in the
LCDM cosmology appear to be the more important factor, rather than the
existence of structure. The OCDM cosmology displays its peak in the
rms values at the highest redshift, and also has a very broad
peak. The TCDM cosmology has both the narrowest range and the lowest
peak. The differences between the SCDM and TCDM cosmologies probably
reflect the differences in structure on different scales in the two
cosmologies, since $R$ is the same for both. In general, however, we
can say that, for all these cosmologies, significant contributions to
the magnification and shear may arise from lenses at a very wide range
of redshifts.
%
%
%

In a similar study, Premadi, Martel and Matzner (1998a) find that the
individual contribution due to each of their lens-planes is greatest
at intermediate redshifts, of order $z=1-2$, for sources located at
$z_s=5$, and Premadi, Martel and Matzner (1998b, c) also find very
broad peaks covering a wide range of intermediate lens-plane redshifts
for sources at $z_s=3$.

\subsection{Results in the different cosmologies}

In Section 5.1, we assumed a value of $\bar{\alpha} = 1$ in the
determination of the angular diameter distances in the various
cosmologies. From the output of the shear algorithm one is able to
obtain an estimate of the clumpiness or smoothness in each
time-slice. Having set the minimum softening scale, the code declares
the number of particles which are assigned the minimum softening, and
one can therefore immediately obtain the mass fraction contained in
clumps, which we choose to define by the minimum softening scale.

For the SCDM cosmology, there is a mass fraction of 0.026 in clumps in
the earliest time-slice at $z=3.6$ (next to $z=3.9$) giving
$\bar{\alpha}(z=3.6)=0.97$, and at $z=0$ the fraction is 0.17, giving
$\bar{\alpha}(z=0)=0.83$. It is clear that the mean value throughout
the redshift range is close to 1, and almost equivalent to the `filled
beam' approximation. (This result is in agreement with Tomita, 1998c,
who finds $\bar{\alpha}$ to be close to 1 in all cases.) The
multiplying factors, $r_dr_{ds}/r_s$, are very close for the values
$\bar{\alpha}=0.83$ and 1.0, although somewhat different from the
values for $\bar{\alpha}=0$ appropriate for an entirely clumpy
universe. The fractional discrepancy between $\bar{\alpha}=0.83$ and
$\bar{\alpha}=1.0$ at the peak of the curves for the SCDM and TCDM
cosmologies is 5.2\% for a source at $z_s = 4$, 2.4\% for $z_s = 2$,
and for sources nearer than $z_s = 1$ the discrepancy is well below
1\%.

In the TCDM cosmology with shape parameter 0.25, $\bar{\alpha}$ falls
to 0.88 at $z = 0$. The higher value in the TCDM model confirms that
the SCDM cosmology (with shape parameter 0.5) has more clumpiness at
late times, allowing higher values of magnification and shear to
occur. The other cosmologies also have high (although somewhat
similar) values for the smoothness parameter at $z = 0$. The values
are 0.80 in the OCDM cosmology, and 0.82 in the LCDM
cosmology. Table~\ref{alphaR} contains the values of
$\bar{\alpha}(z=0)$, which therefore represent the {\em minimum}
values for $\bar{\alpha}$, and also the ratios
$R(\bar{\alpha}=1)/R[\bar{\alpha}(z=0)]$ at the peaks of the curves
for the different cosmologies. These have all been evaluated for a
source redshift of 3.6 to enable direct comparisons to be made.

It has been shown by Barber et al. (1999) that the weak lensing
statistics show only a small sensitivity to the smoothness parameter,
$\bar{\alpha}$, for values between 0.83 and 1 in the SCDM
cosmology. Furthermore, the minimum values in all the cosmologies were
always at least 0.8 throughout the redshift range, so that the real
values are likely to be closer to unity in every case. We have
therefore chosen to present our results on the basis of
$\bar{\alpha}=1$ throughout for all the cosmologies. This conclusion
is validated qualitatively, since the variable softening scheme used
in the algorithm ensures that almost all rays pass entirely through
softened mass.

The source redshifts, $z_s$, we have chosen throughout this work are
close to 4, 3, 2, 1 and 0.5, and we shall refer to the sources in
these terms. The actual redshift values vary slightly for the
different cosmologies and appear in Table~\ref{mu4C}.

\begin{figure}
$$\vbox{
\psfig{figure=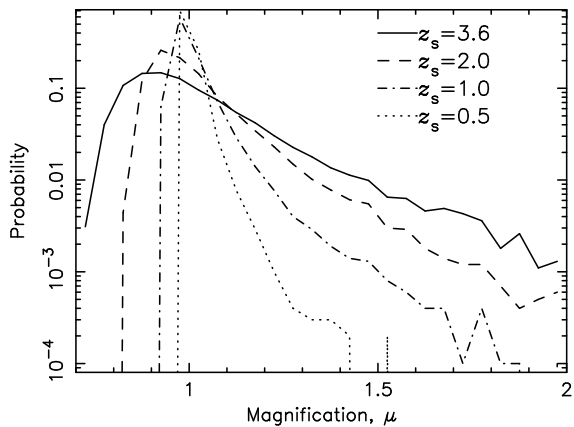,width=8.7truecm,angle=270}
}$$
\caption{Probability distributions for the magnification, for $z_s=3.6,$
2.0, 1.0 and 0.5 in the LCDM cosmology.}
\label{magdistz4aL}
\end{figure}

We show in Figure~\ref{magdistz4aL} an example of the distributions of
the magnifications, $\mu$, in the LCDM cosmology for four different
source redshifts. For all the source redshifts and all the
cosmologies, there is a significant range of magnification. From these
distributions we have computed the values at the peaks,
$\mu_{\mathrm{peak}}$. Then, since the distributions are asymmetrical,
we have calculated the values, $\mu_{\mathrm{low}}$ and
$\mu_{\mathrm{high}}$, above and below which 97\mbox{$\frac{1}{2}$}\%
of all lines of sight fall, and also the rms deviations from unity for
the magnifications. (These latter rms values have been computed only
for the lines of sight displaying magnifications between
$\mu_{\mathrm{low}}$ and $\mu_{\mathrm{high}}$ because three high
magnification events in the LCDM cosmology would distort the rms
values considerably.) All the values mentioned are displayed in
Table~\ref{mu4C}.
%
%
%
\begin{table}
\begin{center}
\begin{tabular}{|c|c|c|c|c|}
\hline
$z_s$ & $\mu_{\mathrm{low}}$ & $\mu_{\mathrm{peak}}$ &
rms deviation & $\mu_{\mathrm{high}}$ \\ \hline
\underline{SCDM} & & & & \\ 
3.9 & 0.835 & 0.933 & 0.115 & 1.420 \\
3.0 & 0.852 & 0.949 & 0.101 & 1.367 \\
1.9 & 0.885 & 0.947 & 0.079 & 1.277 \\
1.0 & 0.930 & 0.973 & 0.049 & 1.181 \\
0.5 & 0.969 & 0.985 & 0.023 & 1.089 \\ \hline
\underline{TCDM} & & & & \\ 
3.9 & 0.861 & 0.959 & 0.091 & 1.315 \\
3.0 & 0.877 & 0.951 & 0.081 & 1.286 \\
1.9 & 0.904 & 0.966 & 0.064 & 1.228 \\
1.0 & 0.941 & 0.972 & 0.039 & 1.144 \\
0.5 & 0.974 & 0.986 & 0.019 & 1.067 \\ \hline
\underline{OCDM} & & & & \\ 
4.0 & 0.858 & 0.919 & 0.115 & 1.469 \\
2.9 & 0.884 & 0.939 & 0.115 & 1.469 \\
2.0 & 0.915 & 0.942 & 0.069 & 1.283 \\
1.0 & 0.960 & 0.972 & 0.033 & 1.147 \\
0.5 & 0.985 & 0.989 & 0.013 & 1.062 \\ \hline
\underline{LCDM} & & & & \\ 
3.6 & 0.789 & 0.885 & 0.191 & 1.850 \\
2.0 & 0.870 & 0.934 & 0.108 & 1.453 \\
1.0 & 0.944 & 0.966 & 0.045 & 1.191 \\
0.5 & 0.981 & 0.987 & 0.016 & 1.070 \\ 
\hline
\end{tabular}
\end{center}
\caption{Various magnification statistics for the different cosmologies, as described in the text.}
\label{mu4C}
\end{table}

\begin{figure}
$$\vbox{
\psfig{figure=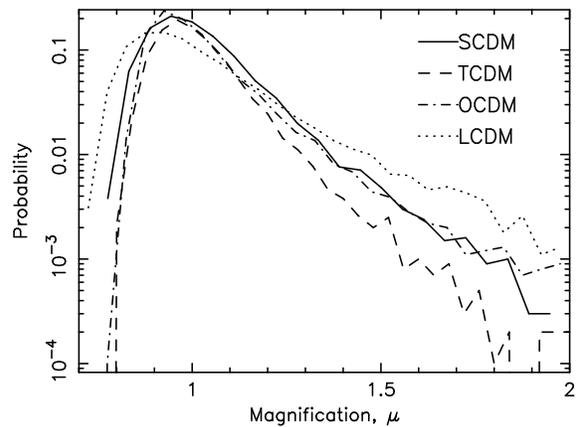,width=8.7truecm,angle=270}
}$$
\caption{The magnification probability distributions for all the cosmologies, assuming $z_s = 4$.} 
\label{magdistz4C}
\end{figure}

\begin{figure}
$$\vbox{
\psfig{figure=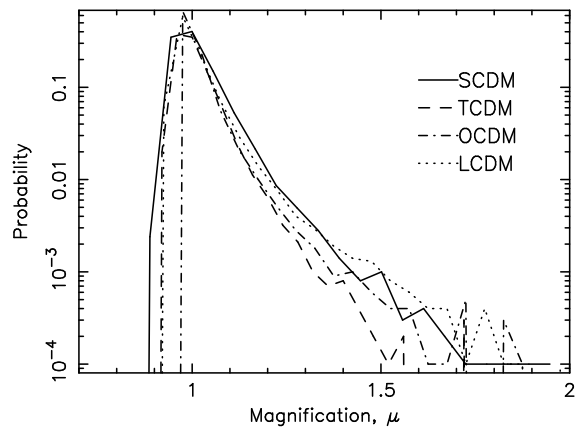,width=8.7truecm,angle=270}
}$$
\caption{The magnification probability distributions for all the cosmologies, assuming $z_s = 1$.} 
\label{magdistz1C}
\end{figure}

It is interesting to compare the distributions in the different
cosmologies. Figure~\ref{magdistz4C} and~\ref{magdistz1C} show the
magnification distributions for all the cosmologies for source
redshifts of 4 and 1, respectively. The distributions for the
magnifications (and also the convergence, shear and ellipticities) are
all broader in the SCDM cosmology when compared with the TCDM
cosmology, due to its more clumpy character. For source redshifts of 4
the OCDM and SCDM cosmologies have very similar distributions even
though the angular diameter distance multiplying factors are larger in
the OCDM cosmology. For high source redshifts the magnification
distributions are broadest in the LCDM cosmology (and the maximum
values of the magnification are greatest here), but for lower source
redshifts the width of the distribution is below the SCDM and OCDM
cosmologies.  
%
%
%
%
%
%

\begin{figure}
$$\vbox{
\psfig{figure=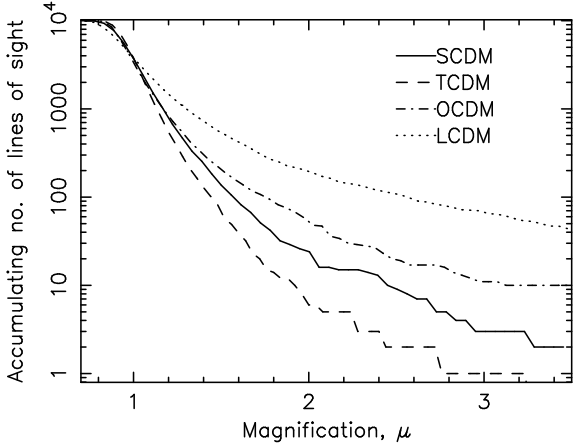,width=8.7truecm,angle=270}
}$$
\caption{Accumulating number of lines of sight for which the magnification is greater than the abscissa value. The plot shows the data for all the cosmologies for $z_s=4$.}
\label{magaccz4C}
\end{figure}

\begin{figure}
$$\vbox{ \psfig{figure=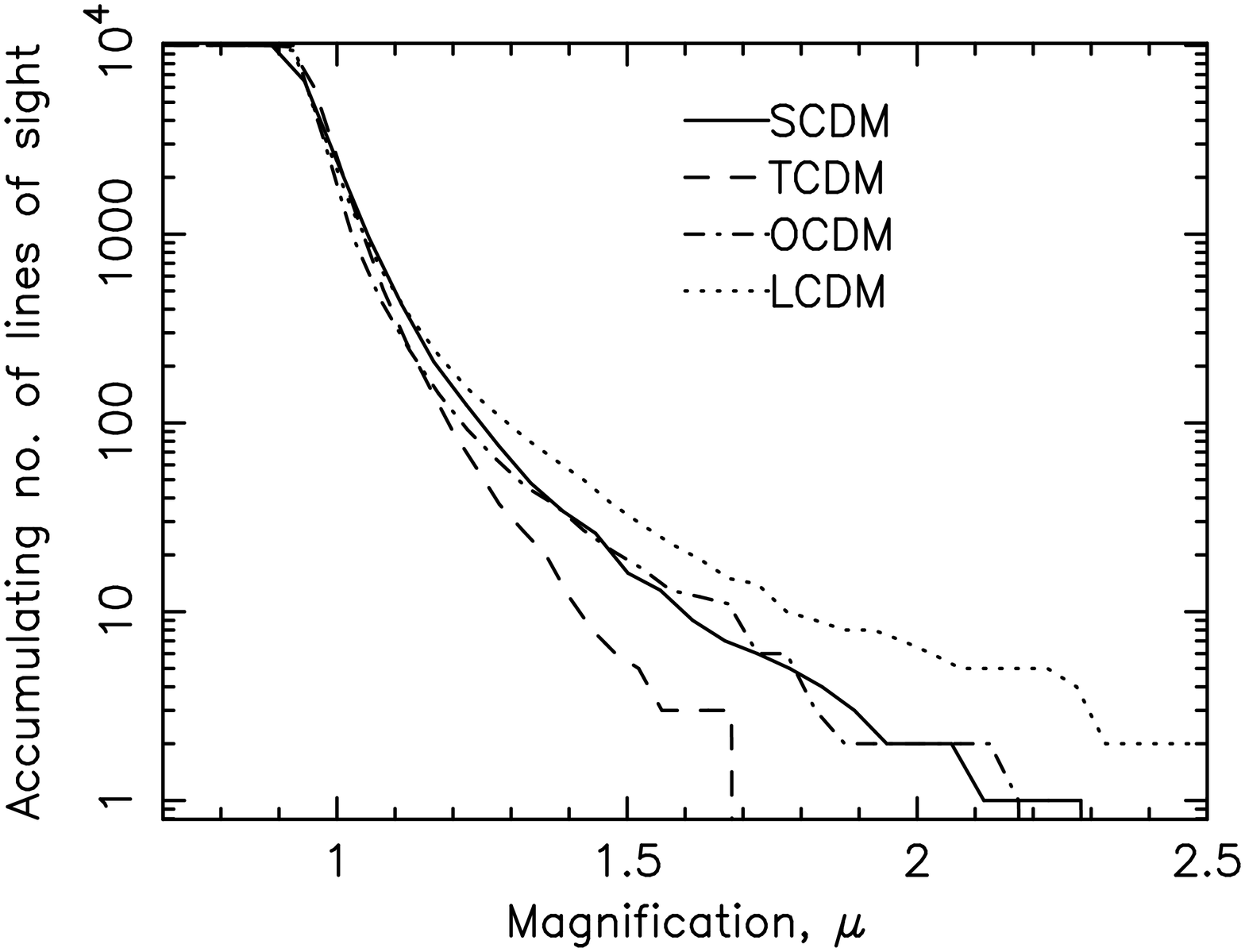,width=8.7truecm,angle=270} }$$
\caption{Accumulating number of lines of sight for which the magnification is greater than the abscissa value. The plot shows the data for all the cosmologies for $z_s=1$.}
\label{magaccz1C}
\end{figure}

\begin{figure}
$$\vbox{
\psfig{figure=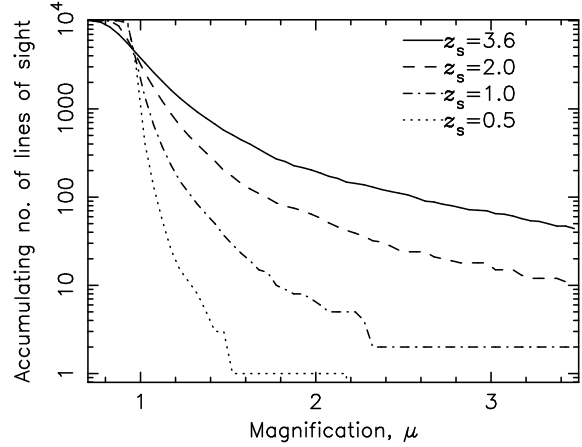,width=8.7truecm,angle=270}
}$$
\caption{Accumulating number of lines of sight for which the magnification is greater than the abscissa value. The plot shows the data for the LCDM cosmology for four different source redshifts as indicated.}
\label{magaccL}
\end{figure}

We plot in Figures~\ref{magaccz4C} and~\ref{magaccz1C} the
accumulating number of lines of sight having magnifications greater
than the abscissa values. This is done for all the cosmologies for
source redshifts of 4 and 1 respectively, and clearly shows the
distinctions at the high magnification end. In particular, the LCDM
cosmology exhibits a very broad tail for $z_s = 4$, and for this
reason we show the accumulating magnifications for this cosmology in
Figure~\ref{magaccL} for source redshifts of 4, 2, 1 and 0.5.
%
%
%
%
%
%
%
%
%

\begin{figure}
$$\vbox{
\psfig{figure=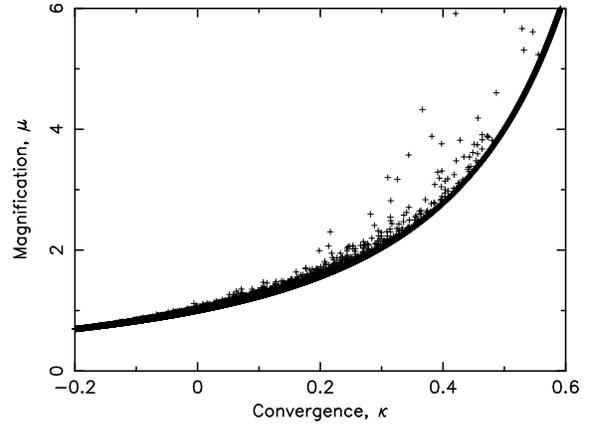,width=8.7truecm,angle=270}
}$$
\caption{LCDM cosmology: $\mu$ vs. $\kappa$ for $z_s=4$ (crosses). The 
continuous line,
shown for comparison, represents $\mu = 1/(1-\kappa)^2$.}
\label{magz4aL}
\end{figure}

In Figure~\ref{magz4aL} we show the magnification, $\mu$, plotted
against the convergence, $\kappa$, for $z_s=4$, again for the LCDM
cosmology. Departures from the curve represented by the values of
$1/(1-\kappa)^2$ clearly arise as a result of the presence of the term
$-\gamma^2$ in the denominator of equation~\ref{musim}, and are most
pronounced at the high $\kappa$ end, as might be expected. This is
true for all the source redshifts and all the cosmologies.
%
%
%

\begin{figure}
$$\vbox{ \psfig{figure=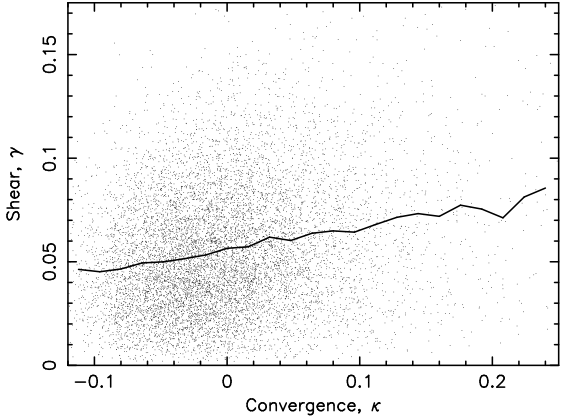,width=8.7truecm,angle=270} }$$
\caption{SCDM cosmology: Shear vs. convergence for sources at 
$z_s=4$ (dots), and the
average shear (full line) in each of the $\kappa$ bins, which shows a
slow and nearly linear increase with increasing convergence.}
\label{onesiggamS}
\end{figure}

We would generally expect the shear, $\gamma$, to fluctuate strongly
for light rays passing through regions of high density (high
convergence), and we indeed find considerable scatter in the shear
when plotted against the convergence. This would result in different
magnification values along lines of sight for which the convergence
values are the same. Figure~\ref{onesiggamS} (reproduced from Barber
et al., 1999) shows the result of binning the convergence values in
the SCDM cosmology and calculating the average shear in each bin, for
sources at $z_s=4$. We see that throughout most of the range in
$\kappa$ the average shear increases very slowly, and closely
linearly, and we have found similar trends in the other
cosmologies. (At the high $\kappa$ end there are too few data points
to establish accurate average values for $\gamma$.) This result
suggests that there may be a trend towards higher mean magnification
values (as $\kappa$ increases) than would be the case if $\langle
\gamma \rangle$ were constant.
%
%
%
%
%
%
%
%
%
%
%
%

\begin{figure}
$$\vbox{
\psfig{figure=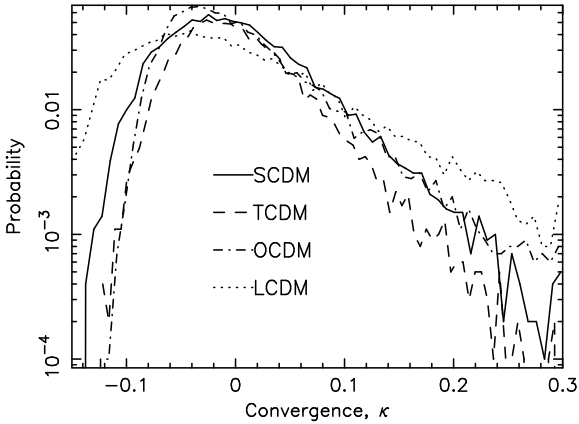,width=8.7truecm,angle=270}
}$$
\caption{The probability distributions for the convergence in the different cosmologies, assuming $z_s = 4$.} 
\label{kappadistz4C}
\end{figure}

\begin{figure}
$$\vbox{
\psfig{figure=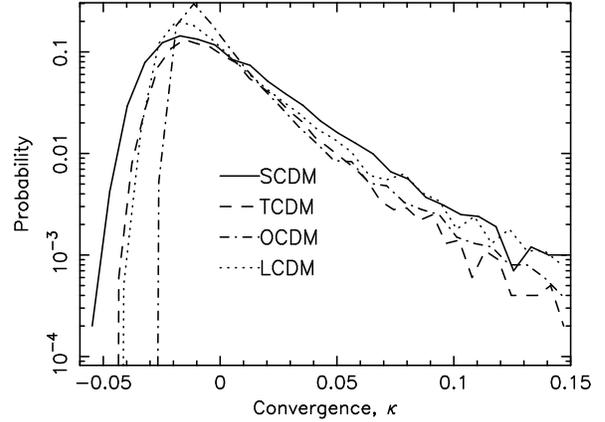,width=8.7truecm,angle=270}
}$$
\caption{The probability distributions for the convergence in the different cosmologies, assuming $z_s = 1$.} 
\label{kappadistz1C}
\end{figure}

Figures~\ref{kappadistz4C} and~\ref{kappadistz1C} show the
distributions in the convergence, $\kappa$, primarily responsible for
the magnifications for all the cosmologies, for $z_s=4$ and $z_s=1$
respectively.
%
%
%
%
%
%

Table~\ref{rmsk} shows the rms values for $\kappa$ in the different
cosmologies for all the source redshifts, and compares them with the
rms values for the magnifications. (This time, unlike
Table~\ref{mu4C}, the rms values have been computed from all lines of
sight, rather than just those with magnifications between
$\mu_{\mathrm{low}}$ and $\mu_{\mathrm{high}}$, so we have not
included the values for source redshifts of $z_s = 3.6$ in the LCDM
cosmology where three strong lensing events occurred.)
\begin{table}
\begin{center}
\begin{tabular}{|c|c|c|c|c|}
\hline
$z_s$ & Magnification & Convergence rms \\
      & rms deviation & rms deviation \\ 
\hline
\underline{SCDM} & & \\ 
3.9 & 0.171 & 0.064 \\
3.0 & 0.149 & 0.058 \\
1.9 & 0.115 & 0.047 \\
1.0 & 0.073 & 0.031 \\
0.5 & 0.037 & 0.016 \\ 
\hline
\underline{TCDM} & & \\ 
3.9 & 0.126 & 0.052 \\
3.0 & 0.111 & 0.047 \\
1.9 & 0.088 & 0.038 \\
1.0 & 0.056 & 0.025 \\
0.5 & 0.027 & 0.013 \\ 
\hline
\underline{OCDM} & & \\
4.0 & 0.245 & 0.066 \\
2.9 & 0.176 & 0.056 \\
2.0 & 0.123 & 0.044 \\
1.0 & 0.060 & 0.024 \\
0.5 & 0.026 & 0.012 \\ 
\hline
\underline{LCDM} & & \\ 
3.6 & N/A & N/A  \\
2.0 & 0.314 & 0.064 \\
1.0 & 0.120 & 0.032 \\
0.5 & 0.030 & 0.013 \\ 
\hline
\end{tabular}
\end{center}
\caption{The rms deviations in the magnification and convergence for the different cosmologies.}
\label{rmsk}
\end{table}

The distributions in the shear, $\gamma$, (defined according to
equation~\ref{gammadef}) are broadest, as expected, for the highest
source redshifts, and, as before, the LCDM cosmology displays the
broadest distribution for $\gamma$ for sources at high redshift.

\begin{figure}
$$\vbox{
\psfig{figure=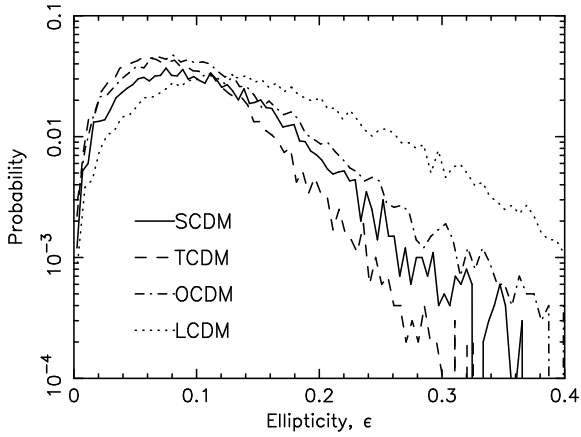,width=8.7truecm,angle=270}
}$$
\caption{The probability distributions for the ellipticity for all the cosmologies, for $z_s = 4$.} 
\label{epsdistz4C}
\end{figure}

\begin{figure}
$$\vbox{
\psfig{figure=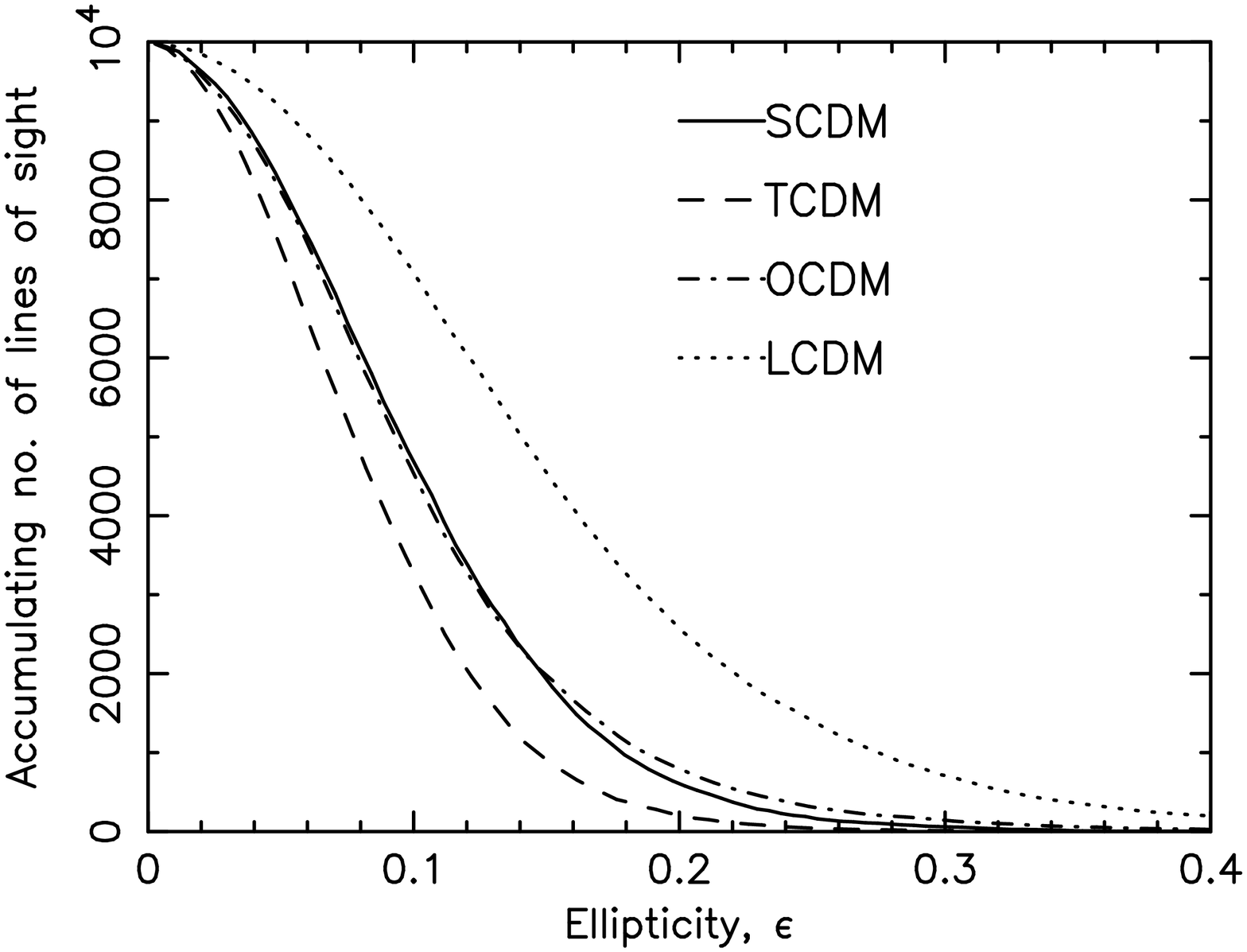,width=8.7truecm,angle=270}
}$$
\caption{The accumulating number of lines of sight for which the ellipticity is greater than the abscissa value. The plot shows the data for all the cosmologies for $z_s = 4$.} 
\label{epsaccz4C}
\end{figure}

\begin{figure}
$$\vbox{
\psfig{figure=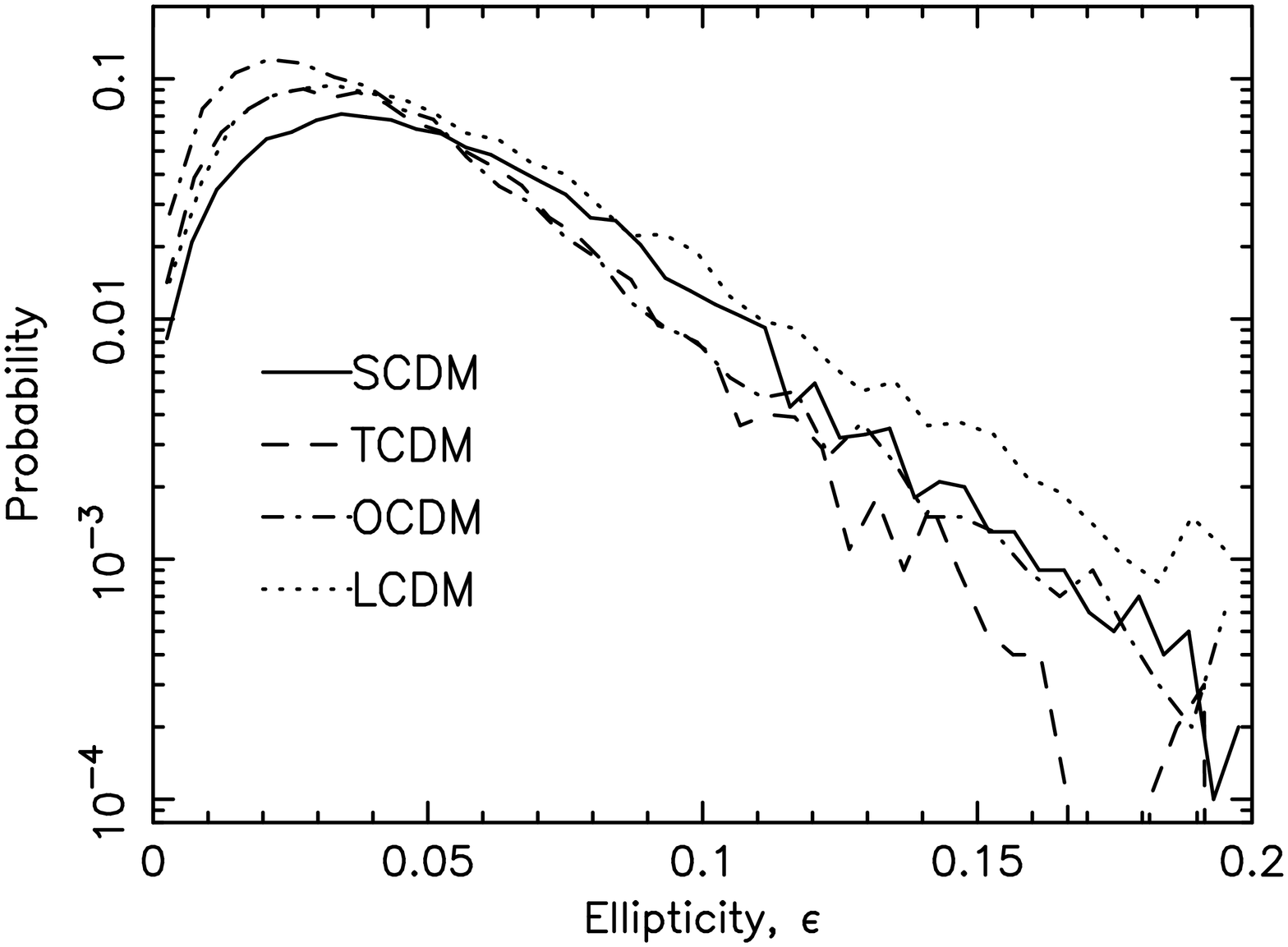,width=8.7truecm,angle=270}
}$$
\caption{The probability distributions for the ellipticity for all the cosmologies, for $z_s = 1$.} 
\label{epsdistz1C}
\end{figure}

\begin{figure}
$$\vbox{
\psfig{figure=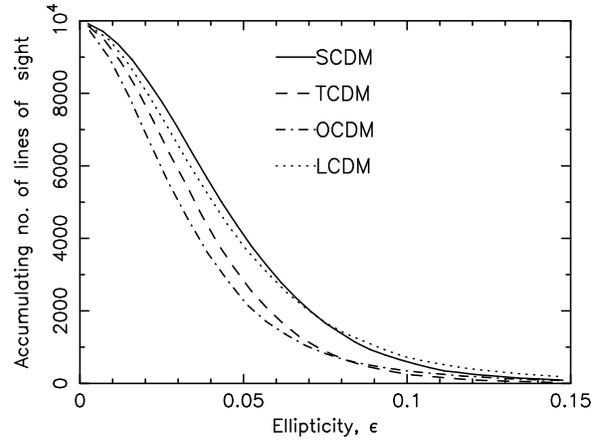,width=8.7truecm,angle=270}
}$$
\caption{The accumulating number of lines of sight for which the ellipticity is greater than the abscissa value. The plot shows the data for all the cosmologies for $z_s = 1$.} 
\label{epsaccz1C}
\end{figure}

The ellipticity, $\epsilon$, in the image of a source is primarily
produced by the shear, and we show in Figure~\ref{epsdistz4C} the
probability distributions for $\epsilon$ for $z_s=4$ for all the
cosmologies. Figure~\ref{epsaccz4C} shows the accumulating number of
lines of sight for which the ellipticity is greater than the abscissa
value, again for $z_s = 4$ and for all the
cosmologies. Figures~\ref{epsdistz1C} and~\ref{epsaccz1C} show the
same information for $z_s = 1$.
%
%
%
%
%
%
%
%
%
%
%
%

For $z_s = 4$, the peaks in the ellipticity distributions occur at
$\epsilon = 0.075$ (SCDM), 0.057 (TCDM), 0.081 (OCDM), and 0.111
(LCDM). For $z_s = 1$, the corresponding figures are $\epsilon =
0.034$ (SCDM), 0.027 (TCDM), 0.021 (OCDM), and 0.033 (LCDM). Once
again, we see that the LCDM cosmology produces the greatest variation
and the highest values for the highest redshifts, although this is not
the case at lower redshifts.

The ellipticity is very closely linear in terms of $\gamma$ throughout
most of the range in $\gamma$, for all the cosmologies. Some scatter
occurs because of the factor containing the convergence, $\kappa$, in
equation~\ref{epssim}.

Finally, the distance-redshift relation, equation~\ref{D1}, implies
that there may be an effect on the angular diameter distances from the
shear. Barber et al. (1999) investigated this for the SCDM cosmology,
and found that whilst the maximum effects of shear on the mean
magnification values may be at least 10\%, only 3.2\% of the lines of
sight were affected in this way. Also, they pointed out that the shear
has an effect in the distance-redshift relation equivalent to
increasing the effective smoothness parameter, $\bar\alpha$, and this
is true for all the cosmologies. By substituting the mean shear value
(from the SCDM cosmology) determined for sources at $z_s=0.5$ (where
the term in $\sigma$ in equation~\ref{D1} is largest), they found the
effect on $\bar\alpha$ (and therefore on the angular diameter
distances) to be completely negligible. Furthermore, the importance of
the effect reduces with redshift, so that our decision to ignore the
effects of shear in the distance-redshift relation is justified.
%
%
%

\section{Determination of the cosmological parameters}
\subsection{Weak shear statistics and the density parameter}

Jain and Seljak (1997) have given careful consideration to the
interpretation of observed shear data (from measured galaxy
ellipticities) by comparing with the analytical results of second
order perturbation theory. They claim that non-linear evolution leads
to non-Gaussian effects in the weak lensing statistics which are more
easily detected in second, and higher order, moments. In particular,
the probability distribution for the three-dimensional density contrast,
\begin{equation}
\delta(\vec{r})=\frac{\rho(\vec{r})}{\bar{\rho}}-1,
\label{delta}
\end{equation}
is almost indistinguishable in the different cosmologies, whereas the
probability distribution function for the convergence, $\kappa$, shows
different peak amplitudes and different dispersions in the different
cosmologies. This is because the transition to non-linearity in the
evolution of structure depends primarily on the density contrast
alone, but the weak lensing signal is strongest in those cosmologies
which have developed structure at optimum redshifts for given source
positions, and therefore depends more directly on the rate of
evolution. Specifically, the non-linear evolution of the power
spectrum introduces non-Gaussianity to the weak lensing statistics,
whilst the matter density parameter, $\Omega_0$, (and to some degree
the vacuum energy density parameter, $\lambda_0$) determines the
dispersion in the statistics.

\begin{figure}
$$\vbox{
\psfig{figure=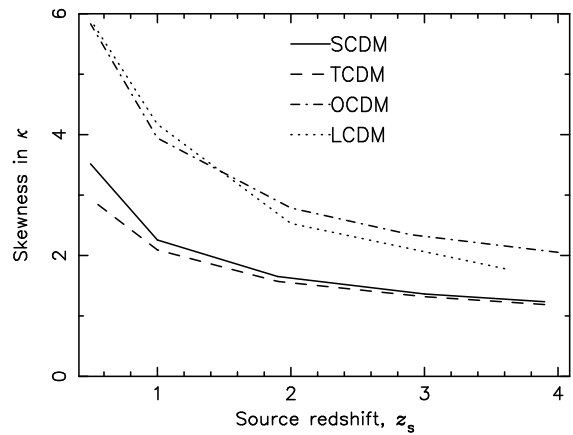,width=8.7truecm,angle=270}
}$$
\caption{The skewness in $\kappa$ as a function of redshift for the different cosmologies, indicating broadly that the skewness decreases with the density parameter, $\Omega_0$.} 
\label{skew1}
\end{figure}

Jain and Seljak (1997) have computed the expected skewness, $S$, in
the convergence,
\begin{equation}
S=\frac{1}{\sigma^3}\langle (\kappa - \bar{\kappa})^3 \rangle,
\label{skew}
\end{equation}
where $\sigma$ represents the standard deviation in
the distribution for $\kappa$. As expected, the skewness was greatest
on small angular scales, and largest for sources at the lowest
redshifts. In general, the LCDM cosmology produced the greatest
skewness, followed by the OCDM, and finally the SCDM cosmologies, for
sources at low redshift. Even though they have assumed an empty beam
scenario with $\bar{\alpha} = 0$, and correspondingly different
angular diameter distances from our work, our own results for the
skewness in $\kappa$ are quite consistent with theirs. We have computed
the skewness, not on different angular scales, but as a function of
redshift for the different cosmologies. It is not possible to state a
specific angular scale for our data, because of the variable softening
approach in the shear algorithm. Figure~\ref{skew1} confirms that, at
redshifts less than about 1.5, the LCDM cosmology gives rise to the
largest skewness, followed by the OCDM cosmology, and finally, the
SCDM and TCDM cosmologies. The difference between the SCDM and TCDM
cosmologies does suggest different forms of structure, which result
directly from the input shape parameter, $\Gamma$, in the power
spectrum. In broad terms, it is clear that the skewness decreases with
source redshift, and decreases with the density parameter, precisely
as expected from perturbation theory.
%
%
%

\begin{figure}
$$\vbox{
\psfig{figure=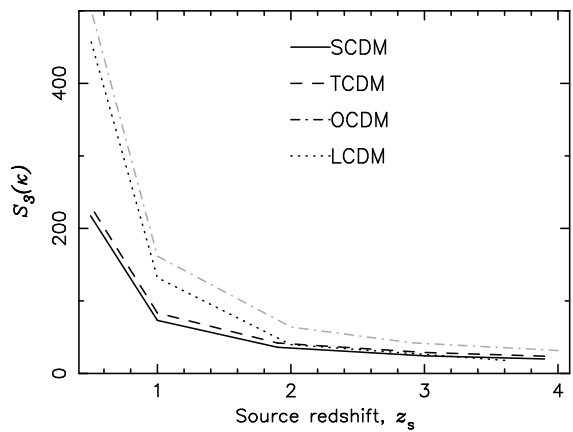,width=8.7truecm,angle=270}
}$$
\caption{The statistic $S_3(\kappa)$, defined by equation~\ref{S3}, as a function of redshift for the different cosmologies. It is clear that the low density cosmologies have higher values of $S_3$ at low redshifts.} 
\label{skew2}
\end{figure}

Bernardeau, van Waerbeke and Mellier (1997) also use perturbation
theory (and the empty beam approach) to assess how the low order
moments in the convergence may depend on the cosmological
parameters. A number of results are predicted. They define the moment,
$S_3$, ({\em not} the skewness, defined above) by
\begin{equation}
S_3 = \frac{\langle \kappa^3 \rangle}{\langle \kappa^2 \rangle ^2},
\label{S3}
\end{equation}
and predict that $S_3 \propto \Omega_0^{-0.8}$ for $z_s \sim 1$, or
$S_3 \propto \Omega_0^{-1.0}$ for $z_s \ll 1$, and they predict a
slightly weaker dependence on $\Omega_0$ for $z_s > 1$. We do not find
these exact relationships because working in the full beam
approximation with variable smoothing alters the predictions. However,
we do find that the value of this statistic does decrease with
increasing $\Omega_0$. Our results for $S_3$ are displayed in
Figure~\ref{skew2}. Bernardeau, van Waerbeke and Mellier (1997)
predict significantly less dependence on $\lambda_0$.
%
%
%

\begin{figure}
$$\vbox{
\psfig{figure=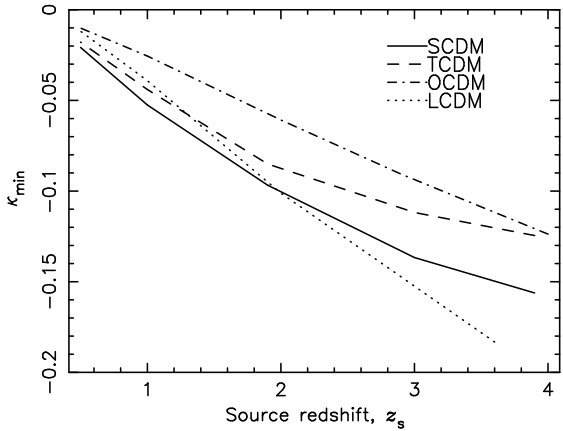,width=8.7truecm,angle=270}
}$$
\caption{$\kappa_{\mathrm{min}}$ vs. redshift for the different cosmologies, showing the order at low redshift expected from the values of $\Omega_0$.} 
\label{skew3}
\end{figure}

\begin{figure}
$$\vbox{ \psfig{figure=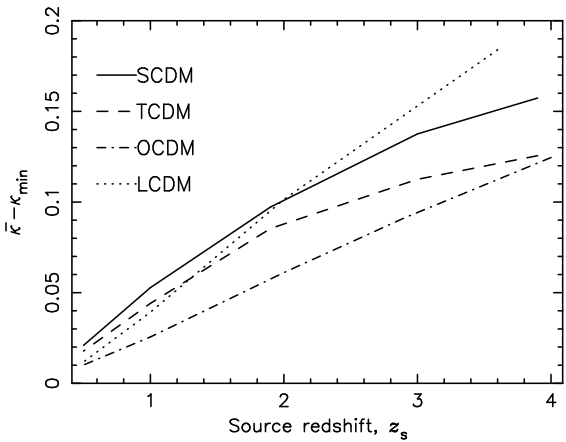,width=8.7truecm,angle=270} }$$
\caption{$(\langle \kappa \rangle -\kappa_{\mathrm{min}})$ vs. redshift for the different cosmologies, showing the order at low redshift expected from the values of $\Omega_0$.} 
\label{skew4}
\end{figure}

Jain, Seljak and White (1999) have used ray-tracing in $N$-body
simulations, as described in the Introduction, in an attempt to
evaluate the density parameter from weak lensing statistics. Since
they make use of the empty beam approximation, they describe the
possibility of many more lines of sight passing through voids from
$z_s = 1$ in open universes, because of the earlier beginning to
structure formation. They suggest that this means that the probability
distribution for $\kappa$ in open universes will have its peak close
to the minimum value. Higher density universes will have the peak well
away from the minimum, so that the shape of the distribution will be
less steep. Moreover, they state that this situation only applies on
small scales, and on such scales this points to the minimum value of
$\kappa$ being roughly proportional to $\Omega_0$. A better statistic
in the empty beam approximation would be that $(\langle \kappa \rangle
-\kappa_{\mathrm{min}}) \propto \Omega_0$. However, they admit that
the result `does depend somewhat on the geometry, as the pathlength
and angular scale differ between open and cosmological constant models
with the same $\Omega_0.$' More specifically, this result will depend
crucially on the approximation used, i.e., the empty or full beam
approximation, and the procedure for particle smoothing. They find
that the measured minimum for $\kappa$ in the open cosmology is close
to the predicted value, but that in the critical, $\Omega_0 = 1$,
cosmology the value was far from the empty beam value because there
were actually no completely empty lines of sight. Our full beam
approach will of course represent the extreme case. It is therefore
reassuring to see that both our $\kappa_{\mathrm{min}}$
(Figure~\ref{skew3}) and $(\langle \kappa \rangle
-\kappa_{\mathrm{min}})$ (Figure~\ref{skew4}) values for the different
cosmologies are in the order expected from the values of
$\Omega_0$. However, because of the entirely different approach from
that of Jain, Seljak and White (1999), the results are not able to
indicate specific values for $\Omega_0$.
%
%
%
%
%
%

Jain, Seljak and White (1999) also find for the $S_3$ statistic that
$S_3(\mathrm{OCDM})$ $>$ $S_3(\mathrm{LCDM})$ $>$ $S_3(\mathrm{TCDM})$
$>$ $S_3(\mathrm{SCDM})$ at low redshift. This concurs with our
findings (see Figure~\ref{skew2}). They quantify the difference
between the SCDM and TCDM cosmologies in terms of a weak function of
the shape parameter, which occurs in the expression for $S_3$ from
perturbation theory.

They have also explained in some detail how observational data may be
used to reconstruct the convergence, which is only feasible with large
fields where the weak shear signal is measurable. However, in view of
our quite different approaches, which clearly give rise to
discrepancies (although only in terms of degree), care must be taken
when interpreting observational data in this way. If we had a better
understanding of the form, distribution and evolution of the dark and
luminous matter in the universe, it might be possible to produce
simulations and weak lensing experiments in more realistic scenarios.

\subsection{Weak lensing of high-redshift Type Ia Supernov\ae}

We have seen in Section 5 the significant ranges in magnifications
(dependent on the cosmology) which might apply to distant sources. In
the absence of magnification (or demagnification) from the large scale
structure, it would be possible to determine the cosmological
parameters, $\Omega_0$ and $\lambda_0$, from the departures from
linearity in the Hubble diagram, provided `standard candle' sources
together with good calibration were available for measurement at high
redshift. This is precisely the route taken by a number of authors,
most importantly Riess et al. (1998) and Perlmutter et al. (1999),
both of whom have used high-redshift Type Ia Supernov\ae~data of
redshifts up to 0.97. It is evident from our results that full account
must be taken of the ranges in magnification for each of the
cosmologies, and in particular the cosmologies suggested by the
high-redshift Type Ia Supernov\ae~results.

Both groups of workers, i.e., Riess et al. (1998) and Perlmutter et
al.  (1999), point to cosmologies which are close to the $\Omega_0 =
0.3,$ $\lambda_0 = 0.7$ cosmological simulation we have analysed in
terms of weak lensing. Consequently, our results from this cosmology
are of considerable interest for their impact on the determination of
the cosmological parameters. Using the data we report here, Barber
(2000) finds that true underlying cosmologies having a deceleration
parameter $q_0 = -0.51 +0.03/-0.24$ may be interpreted as having $q_0
= -0.55$, from the use of perfect standard candles (without intrinsic
dispersion), arising purely from the effects of weak lensing. This
significant dispersion in $q_0$ (approximately $2\sigma$) is somewhat
larger than that found by Wambsganss et al. (1997) based on a
cosmology with $\Omega_M = 0.4,$ $\Omega_{\Lambda} = 0.6$, because of
our broader magnification distribution at $z = 1$.

\section{SUMMARY AND CONCLUSIONS}

In the application of the code for the three-dimensional shear, we
have had to consider what appropriate angular diameter distance values
should be applied to the data. The values have all been calculated
numerically from the generalised beam equations (see section 4) for
the different cosmologies. This was done in each case for a smoothness
parameter $\bar{\alpha} = 1$. Our variable softening scheme for the
particles ensures that nearly all rays pass entirely through softened
mass, and in addition, we found that the minimum value of
$\bar{\alpha}$ was at least 0.8 (at $z = 0$) in all the cosmologies.
The differences in the magnification distributions for $\bar{\alpha} =
1$ and for the minimum value were almost indistinguishable. The
effects of shear on the angular diameter distances (through changes to
the effective value of the smoothness parameter, $\bar\alpha$), were
found to be completely negligible at all redshifts, so we are
justified in ignoring them in the distance-redshift
relation. Furthermore, they are always additive, making the effective
value of $\bar{\alpha}$ even closer to unity.

We found, in Section 5.1, that the Einstein-de Sitter universes showed
the most rapid growth in the `intrinsic' shear components at late
times, as expected from the growth of structure in these cosmologies.
This was as a result of studying the computed shear values before the
application of the angular diameter distance factors, and before
conversion to physical units.  The OCDM and LCDM cosmologies indicated
only a limited contribution from the cosmological constant in terms of
the growth in the shear values, again consistent with the expected
evolution of structure in these cosmologies.

When the computed shear values were multiplied by the full conversion
factors appropriate to the integration, together with the angular
diameter distance factors, the resulting curves exhibited very broad
peaks, indicating that significant lensing may result from structure
in a wide band of redshifts. Significantly, the LCDM cosmology has
both the broadest and the highest peak, indicating that this cosmology
should produce, for example, the broadest range of
magnifications. This result would appear to come primarily from the
large values of the angular diameter distance factors for this
cosmology, rather than any considerations about the evolution of
structure. However, the broader and higher peak for the SCDM
cosmology, compared with the TCDM cosmology, does indicate the
differences in structure within them, since they both have the same
values for the angular diameter multiplying factor, $R$.

In Section 5.2, we showed the results for the magnification
distributions for the different cosmologies for different source
redshifts, and these are concisely summarised in Table~\ref{mu4C}. At
high redshift, the LCDM cosmology produces the highest magnifications,
the broadest distribution curves, and the lowest peak values. For
sources at $z_s=3.6$ in the LCDM cosmology, 97\mbox{$\frac{1}{2}$}\%
of all lines of sight have magnification values up to 1.850. (The
maximum magnifications, not quoted here, depend on the choice of the
minimum softening in the code, although the overall distributions are
very insensitive to the softening.) The rms fluctuations in the
magnification (about the mean) were as much as 0.191 in this
cosmology, for sources at $z_s=3.6$.  Even for sources at $z_s=0.5$
there is a measurable range of magnifications in all the cosmologies.

The immediate implication of these results is the likely existence of
a bias in observed magnitudes of distant objects, and a likely
dispersion for standard candles (for example, Type Ia Supernov\ae) at
high redshift.

The magnification versus the convergence showed the presence of
significant shear, and the mean values of $\gamma$ in small
convergence bins pointed to a possible slow linear increase in
$\langle \gamma \rangle$ with $\kappa$. This would result in a trend
towards higher mean magnification values (as $\kappa$ increases) than
would be the case for constant $\langle \gamma \rangle$.

The distributions for the shear, $\gamma$, are broadest, in the LCDM
cosmology for high source redshifts, and there are closely linear
relationships between the ellipticity and the shear in all the
cosmologies. This relationship leads again to broad distributions in
the ellipticity for high-redshift sources in the LCDM cosmology. The
peak in the ellipticity distribution for the LCDM cosmology, for $z_s
= 3.6$, is 0.111, being almost twice the value in the TCDM cosmology.

Jain, Seljak and White (1999) have expressed the possibility of
determining the value of the density parameter from the convergence
field from weak lensing statistics. They show one-point distribution
functions for $\kappa$, assuming sources at $z_s = 1$, for all their
four cosmologies and using different (fixed) smoothing scales in
each. They describe the increasing non-Gaussianity of the distribution
functions as the smoothing scale is reduced, and the increasing tail
at high $\kappa$. They also describe the shape of the distribution
functions for negative $\kappa$, which results from the rate of
structure formation in the different cosmologies, and claim the
interesting conclusion that the minimum value of $\kappa$ is
proportional to the density parameter. We were able to establish from
our own work that, in broad terms, the skewness in $\kappa$ decreases
with source redshift, and decreases with increasing density parameter,
as expected. Also, the statistic $S_3$ (defined in equation~\ref{S3})
was found to decrease with increasing $\Omega_0$. However, we were
unable to establish more precise relationships, because of the use of
the full beam approximation. The order of the cosmologies for $S_3$ is
the same as that presented by Jain, Seljak and White (1999). We also
found that both $\kappa_{\mathrm{min}}$ and $(\langle \kappa \rangle
-\kappa_{\mathrm{min}})$ are in the correct order for the different
cosmologies in terms of $\Omega_0$, although, again for the same
reason, it would not be possible to obtain specific values for
$\Omega_0$.

We now, very briefly, make some comparisons with the weak lensing
results obtained by other authors. This will not be exhaustive,
because it has been anticipated that our results should be different,
for a number of reasons. Primary amongst these are the following.
First, our results were obtained using the three-dimensional shear
code, which allows periodicity, the use of the peculiar potential, the
net zero mean density requirement, and angular diameter distances to
every evaluation position within each simulation volume. Second, many
two-dimensional (planar) approaches may suffer from inadequate
convergence to the true limiting values for the shear matrix and
angular deflections. Third, we have introduced a physically realistic
variable softening to the method, which requires use of the full beam
approximation for the angular diameter distances, rather than the
empty cone approximation used by many authors with either point
masses, small (fixed) softenings, or small pixellation in the planes. 

The magnification distributions of Jaroszy\'nski et al. (1990) for the
SCDM cosmology do not have mean values of 1, and their dispersions in
the convergence for sources at $z_s=1$ and $z_s=3$ are considerably
lower than our values, with very little evolution with redshift.  In
our work, the peaks of the ellipticity distributions occured at values
of 0.075 (similar to the value for $z_s=4$) and 0.034 for sources at
$z_s =3$ and 1 respectively, and these are somewhat lower than the
values of 0.095 ($z_s=3$) and 0.045 ($z_s=1$) found by Jaroszy\'nski
et al. (1990). Rather surprisingly, however, their peak values in the
distributions for the shear are quite similar to ours, especially for
sources at $z_s=3$.

Wambsganss, Cen and Ostriker (1998) find magnifications up to 100 in
the SCDM cosmology, and correspondingly highly dispersed
distributions, very much larger than ours for $z_s=3$.  The high
magnification tail in the distributions almost certainly derives from
the low value of the (fixed) softening scale resulting from the
`smearing' of the mass distribution in the $10h^{-1}$kpc $\times
10h^{-1}$kpc pixels.

Similarly, Marri and Ferrara (1998) show wide magnification
distributions, and very high maximum values, again which occur as a
result of using point particles rather than smoothed particles. Their
procedure (summarised in Section 1.2) is completely different from
ours, and, unlike us, they find that the SCDM cosmology has the
broadest magnification probability distribution, followed by the LCDM
cosmology, and finally their HCDM cosmology. In particular, we would
disagree with their choice of $\bar\alpha = 0$, which is
representative of an entirely clumpy universe, as opposed to our
finding, that the SCDM universe is close to being smooth at all
epochs.

Hamana, Martel and Futamase (1999) find that the dispersions in the
probability distributions for $\kappa$, $\gamma$ and $\mu$ are all
greatest for the Einstein-de Sitter cosmology, and are very similar
for their open and cosmological constant cosmologies. This is in
complete contrast to our findings. Their magnification and convergence
distributions for $z_s = 1,$ 2 and 3 are much broader than ours,
although the distributions for $\gamma$ in the SCDM cosmology are
similar.

The ranges in magnification from Premadi, Martel and Matzner (1998a)
appear to be rather similar to ours for sources at $z_s=3$ for their
three cosmologies, and the widths of the distributions are in the same
order for the different cosmologies that we find. This is reassuring
because, although their method relies on two-dimensional projections
of the simulation boxes, they include many of the essential features
to which we have drawn attention, for example, an assumed periodicity
in the matter distribution, randomly chosen initial conditions to
avoid structure correlations between adjacent simulation boxes, the
net zero mean density requirement, realistic mass profiles for the
particles, and use of the filled beam approximation with a smoothness
parameter, $\bar\alpha =1$. They show the average shear for a source
at $z_s=5$ contributed by each of the lens-planes individually, and
find that the largest contributions come from those planes at
intermediate redshift, of order $z$=1 -- 2. Similarly, they find that
the lens-planes which contribute most to the average magnifications
are also located at intermediate redshifts. In terms of the
development of structure in the different cosmologies, they find that
the lens-planes contributing the most shear and magnification are
located at larger redshifts for those cosmologies with smaller
$\Omega_M$. The average shear for each redshift has been plotted by
the authors. They find, for all redshifts, that the values in the SCDM
cosmology are many times greater than in the LCDM cosmology, which is,
in turn, greater than in the OCDM cosmology. This is not what we find.

Fluke, Webster and Mortlock (2000), in applying the ray bundle method
of Fluke, Webster and Mortlock (1999), explain clearly the differences
between the empty cone and full beam approximations. They use the
empty cone approximation, because of their use of an effective fixed
physical radius for each particle, equal to $\sqrt{2}\times$ the
Einstein radius for each. This gives rise to magnification probability
distributions with $\mu_{\mathrm{min}} = 1$, arising from the use of
$\bar{\alpha}=0$, and high magnification tails, arising from the small
effective radii for particles and clusters. They obtain the weak
lensing statistics for the same cosmologies we have used. For $z_s =
1$, the distribution in the magnifications for the SCDM cosmology is
clearly broader than that for the LCDM cosmology, which in turn is
broader than the OCDM cosmology. This is also true of our data,
although the order for the cosmologies, in our work, is completely
altered for higher source redshifts. The authors find, as do we, that
the most significant differences amongst the cosmologies arise
directly as a result of the optical depth to the source, which is
related to the angular diameter distances. Moreover, there is a
suggestion in their results (Fluke, 1999) that Kolmogorov-Smirnov
tests may be unable to distinguish the cosmologies, {\em even random
distributions of particles,} in terms of their magnification
distributions, if the same set of angular diameter distance factors is
applied to each.

This last statement may prove to be extremely important, as it appears
from both our own work and that of Fluke, Webster and Mortlock (2000),
that the angular diameter distances (or optical depths) are really the
determining factor for weak lensing statistics. Moreover, our two
approaches may represent `limits' for the true lensing behaviour from
the mass distribution in the universe. The variable softening facility
within our algorithm leads naturally to the assumption that the
universe may be described in terms of the full beam
approximation. This is, however, quite different from the assumptions
of most other workers, who frequently use point particles, or a
limited form of fixed softening, or small pixellation, and therefore
use the empty beam approximation. The two approaches give rise to
quite different expectations and results. The most obvious differences
are the following. First, strong lensing can occur with effectively
small particles, leading to high magnification tails in the
probability distributions. Second, magnification distributions in the
empty cone approximation all have $\mu_{\mathrm {min}} = 1$, whilst in
the full beam approximation, $\mu_{\mathrm {min}} \leq 1$; this may
alter the dispersions in the two distributions. Third, the mean values
for the magnifications can be calculated from the respective angular
diameter distances in the different cosmologies for the empty beam
approximation; however, the mean values in the full beam approximation
are always 1. These points make comparisons between methods using the
different approximations difficult. However, it is probable that the
universe is neither completely smooth nor filled with galactic-mass
point-like objects. (Subramanian, Cen and Ostriker, 1999, highlight
our uncertainty in this area, by suggesting that small dense masses
formed early during hierarchical clustering may persist to late times,
so that real cluster halo structures may depend crucially on the
detailed dynamics of the dense pockets.) The resolution of this
question together with a better understanding of the form,
distribution and evolution of the dark and luminous matter content of
the universe should provide a much clearer indication of the likely
weak lensing statistics in cosmological $N$-body simulations of the
future.

\section*{ACKNOWLEDGMENTS}

We are indebted to the Starlink minor node at the University of Sussex
for the preparation of this paper, and to the University of Sussex for
the sponsorship of AJB.  PAT is a PPARC Lecturer Fellow. We thank NATO
for the award of a Collaborative Research Grant (CRG
970081). R. L. Webster of Melbourne University has been particularly
helpful.

\section*{REFERENCES}

\baselineskip 0.41cm 
\begin{trivlist} 

\item{Barber, 2000, in preparation}
\item{Barber, A. J., Thomas P. A. and Couchman H. M. P., 1999, MNRAS, in press (astro-ph 9901143).}
\item{Bartelmann M., Ehlers J. \& Schneider P., 1993, A. and A., 280, 351.}
\item{Bernardeau F., van Waerbeke L. \& Mellier Y., A. and A., 1997, 322, 1.}
\item{Carroll, S. M., Press, W. H. and Turner, E. L., 1992, A.R.A.\&A., 30, 499.}
\item{Couchman H. M. P., Barber A. J. and Thomas P. A., 1999, MNRAS, 308, 180.}
\item{Couchman H. M. P., Thomas, P. A., and Pearce F. R., 1995, 
Ap. J., 452, 797.}
\item{Dyer C. C. and Roeder R. C., 1973, Ap. J., 180, L31.} 
\item{Fluke C. J., 1999, private communication.}
\item{Fluke C. J., Webster R. L. and Mortlock
D. J., 1999, MNRAS, 306, 567.}
\item{Fluke C. J., Webster R. L. and Mortlock
D. J., 2000, MNRAS (submitted).}
\item{Hamana T., Martel H. and Futamase T., 1999, astro-ph 9903002, preprint.}
\item{Hockney R. W. and Eastwood J. W., 1988, `Computer Simulation
Using Particles,' IOP Publishing, ISBN 0-85274-392-0.}
\item{Jain B. and Seljak U., Ap. J., 1997, 484, 560.}
\item{Jain B., Seljak U. and White
S., 1998, astro-ph, 9804238, preprint.}
\item{Jain B., Seljak U. and White
S., 1999, astro-ph, 9901191, preprint.} 
\item{Jaroszy\'nski M., Park C., Paczynski B., and Gott III J. R., 
1990, Ap. J., 365, 22.}
\item{Lacey C. and Cole S., 1993, MNRAS, 262, 627.}
\item{Lacey C. and Cole S., 1994, MNRAS, 271, 676.}
\item{Linder E. V., 1998a, A. and A., 206, 175.}
\item{Linder E. V., 1998b, A. and A., 206, 190.}
\item{Marri S. and Ferrara A., 1998, Ap. J., 509, 43.}
\item{Peacock J. A. and Dodds S. J., 1994, MNRAS, 267, 1020.}
\item{Peebles P.J.E., 1993, `Principles of Physical Cosmology,' 
Princeton University Press, ISBN 0-691-07428-3.}
\item{Pei Y. C., 1993, Ap. J., 403, 7.}
\item{Perlmutter S., Aldering G., Goldhaber G., Knop R. A., Nugent P.,
Castro P. G., Deustua S., Fabbro S., Goobar A., Groom D. E., Hook
I. M., Kim A. G., Kim M. Y., Lee J. C., Nunes N. J., Pain R.,
Pennypacker C. R., Quimby R., Lidman C., Ellis R. S., Irwin M.,
McMahon R. G., Ruiz-Lapuente P., Walton N., Schaefer B., Boyle B. J.,
Filippenko A.  V., Matheson T., Fruchter A. S., Panagia N., Newberg H.
J. M., Couch W. J., and Project T. S. C., 1999, Ap. J., 517, 565.}
\item{Premadi P., Martel H. and Matzner R., 1998a, Ap. J., 493, 10.}
\item{Premadi P., Martel H. and Matzner R., 1998b, astro-ph, 
9807127, preprint.}
\item{Premadi P., Martel H. and Matzner R., 1998c, astro-ph, 
9807129, preprint.}
\item{Rauch K. P., 1991, Ap. J., 374, 83.}
\item{Refsdal S., 1970, Ap. J., 159, 357.}
\item{Richstone D. O., Loeb A. and Turner E. L., 1992, Ap. J., 393, 477.}
\item{Riess A. G., Filippenko A. V., Challis P., Clocchiatti A.,
Diercks A., Garnavich P. M., Gilliland R. L., Hogan C. J., Jha S.,
Kirshner R. P., Leibundgut B., Phillips M. M., Reiss D., Schmidt
B. P., Schommer R. A., Smith R. C., Spyromilio J., Stubbs C., Suntzeff
N. B. and Tonry J., 1998, A. J., 116, 1009.}
\item{Schneider P., Ehlers J., and Falco E. E., 1992, `Gravitational 
Lenses,' Springer-Verlag, ISBN 0-387-97070-3.}
\item{Schneider P. and Weiss A., 1988a, Ap. J., 327, 526.}
\item{Schneider P. and Weiss A., 1988b, Ap. J., 330, 1.}
\item{Subramanian K, Cen R. and Ostriker J. P., 1999, astro-ph 9909279, preprint.}
\item{Tomita K., 1998a, Prog. Th. Phys., 99, 97.}
\item{Tomita K., 1998b, astro-ph, 9806003, preprint.}
\item{Tomita K., 1998c, Prog. Th. Phys., 100, 79.}
\item{Viana P. T. P., and Liddle A. R., 1996, MNRAS, 281, 323.}
\item{Wambsganss J., Cen R., and Ostriker J., 1998, Ap. J., 494, 29.}
\end{trivlist}
\begin{description} 
\item 
\end{description} 

\onecolumn

\setcounter{page}{24}

\appendix
\section{GENERALISATION OF THE DYER-ROEDER EQUATION}

In Section 4, we presented the most simple form for the Dyer-Roeder
equation (equation~\ref{DR}), which can be solved analytically for
$\Omega_0 = 1$, $\lambda_0 = 0$, and arbitrary $\bar{\alpha}$. The
general solution for the angular diameter distance between redshifts
of $z_1$ and $z_2$, in such a cosmology, is well documented (see, e.g., Schneider et al., 1992) and from this solution the multiplying factors
$D_dD_{ds}/D_s$ are easily obtained. However, this solution applies
only to cosmologies with zero cosmological constant. We therefore
generalised the form of the Dyer-Roeder equation to apply to the
cosmologies being studied in this work. The following summary of this
work is likely to be helpful to others working in this
field.

We started from the generalised beam equation, quoted by
Linder (1998a and b):
\begin{equation}
\frac{d^2D}{dz^2}+\left[3+q(z)\right](1+z)^{-1}\frac{dD}{dz}+\frac{3}{2}(1+z)^{-2}D\sum_s(1+s)\bar{\alpha}_s(z)\Omega_s(z) = 0.
\label{genbeam}
\end{equation}
In this equation,
\begin{equation}
q(z) = \frac{\frac{1}{2}\sum_s\Omega_s(0)(1+3s)(1+z)^{1+3s}}{\sum_s\Omega_s(0)(1+z)^{1+3s}-\left[\sum_s\Omega_s(0)-1\right]}
\label{q0}
\end{equation}
is the deceleration parameter at redshift $z$, each value of $s$
denotes a content component of the universe (for example,
non-relativistic matter, radiation, vacuum energy, etc.), and
$\bar{\alpha}_s$ and $\Omega_s(z)$ represent the smoothness parameter
and the density parameter respectively, applicable to the component
$s$, and redshift, $z$. For two-component cosmologies, to which we
have restricted our work, $s=0$ for dust and $s=-1$ for the vacuum
energy. When $s=0$ only, equation~\ref{genbeam} reduces to the
Dyer-Roeder equation immediately. Also, with two components only, the
equation for the deceleration parameter at the present day, reduces
from equation~\ref{q0} to the familiar form, $q_0 = \frac{1}{2}\Omega
(0)-\lambda (0)$, where now we have used $\Omega$ and $\lambda$ to
represent the matter and vacuum energy density parameters
respectively. Also, the Hubble parameter is, in general,
\begin{equation}
H(z) = H_0\left\{\sum_s\Omega_s(0)(1+z)^{3(1+s)}-\left[\sum_s\Omega_s(0)-1\right](1+z)^2\right\}^{1/2}.
\label{Hz}
\end{equation}
For a two-component universe this becomes:
\begin{equation}
H(z) = H_0\left\{\Omega(0)(1+z)^3+\lambda(0)-\left[\Omega(0)+\lambda(0)-1\right](1+z)^2\right\}^{1/2}.
\label{Hz2}
\end{equation}

For two components only, the generalised beam equation
(equation~\ref{genbeam}) is:
\begin{equation}
\frac{d^2D}{dz^2}+\left[3+q(z)\right](1+z)^{-1}\frac{dD}{dz}+\frac{3}{2}(1+z)^{-2}D\bar{\alpha}(z)\Omega(z) = 0,
\label{genbeam2}
\end{equation}
in which
\begin{equation}
q(z) = \frac{\frac{1}{2}\left[\Omega(0)(1+z)-2\lambda(0)(1+z)^{-2}\right]}{\Omega(0)z+\lambda(0)(1+z)^{-2}-\lambda(0)+1},
\label{qz2}
\end{equation}
\begin{equation}
\Omega(z) = \Omega(0)(1+z)^3\left[H(z)/H_0\right]^{-2},
\label{Omegaz2}
\end{equation}
\begin{equation}
\frac{H(z)}{H_0} = \left\{\Omega(0)(1+z)^3+\lambda(0)-\left[\Omega(0)+\lambda(0)-1\right](1+z)^2\right\}^{1/2},
\label{HzH0}
\end{equation}
and so
\begin{equation}
\Omega(z) = \Omega(0)(1+z)^3\left\{\Omega(0)(1+z)^3+\lambda(0)-\left[\Omega(0)+\lambda(0)-1\right](1+z)^2\right\}^{-1}.
\label{Omegaz22}
\end{equation}
To solve equation~\ref{genbeam2}, boundary conditions
\begin{equation}
D(z_1,z_1) = 0,
\label{Dz1z1}
\end{equation}
and
\begin{equation}
\frac{dD(z_1,z)}{dz}\mid _{z=z_1} = (1+z_1)\left[\frac{H(z_1)}{H_0}\right]^{-1}
\label{bdy2}
\end{equation}
are set, where the second condition is made by considering the form of
the Hubble law for a fictitious observer at the redshift $z_1$.

However, we need values for the angular diameter distances between any
arbitrary redshift values (not always based on $z=0$) in order to
construct values for $D_dD_{ds}/D_s$ at all the required evaluation
positions. To do this, equation~\ref{genbeam2} has to be generalised
further to apply to any arbitrary redshift, and we can do this by
changing the variable to
\begin{equation}
w \equiv \frac{1+z}{1+z_1}-1.
\label{w}
\end{equation}
$w$ then corresponds to the redshift of an object as if viewed by an
observer at the arbitrary redshift $z_1$. Then substituting the
expression for $q(z)$ (equation~\ref{qz2}), equation~\ref{genbeam2}
becomes, after some manipulation,
\begin{eqnarray}
& &\frac{d^2D}{dw^2}(1+z_1)^2+\left\{3+\frac{\frac{1}{2}\left[\Omega(0)x-2\lambda(0)x^{-2}\right]}{\Omega(0)(x-1)+\lambda(0)x^{-2}-\lambda(0)+1}\right\}\frac{1}{x}\frac{dD}{dw}(1+z_1) \nonumber \\
& &\hskip 1.0 in +\frac{3}{2}D\bar{\alpha}\left\{\frac{\Omega(0)x}{\Omega(0)x^3+\lambda(0)-\left[\Omega(0)+\lambda(0)-1\right]x^2}\right\} = 0,
\label{13new}
\end{eqnarray}
with boundary conditions,
\begin{equation}
D(z_1,z_1) = 0,
\label{Dz1z12}
\end{equation}
and
\begin{equation}
\frac{dD(z_1,w)}{dw}\mid _{w=z_1}=(1+z_1)^{-1}\left\{\Omega(0)(1+z_1)^3+\lambda(0)-\left[\Omega(0)+\lambda(0)-1\right](1+z_1)^2\right\}^{-1/2}.
\label{bdy22}
\end{equation}
(In equation~\ref{13new} we have written $x \equiv (1+w)/(1+z_1)$ for clarity.)

We have solved this equation numerically for all the cosmologies,
checking carefully that the results are the same as the analytical
values for the Einstien-de Sitter model. To solve it, we
made further definitions to simplify the form of the equation. First,
for clarity, we directly interchanged $w$ and $z$, and made the
following definitions.
\begin{equation}
a \equiv (1+z_1)^2,
\label{a}
\end{equation}
\begin{equation}
b \equiv \Omega(0)/(1+z_1),
\label{b}
\end{equation}
\begin{equation}
c \equiv 2\lambda(0)(1+z_1)^2,
\label{c}
\end{equation}
\begin{equation}
d \equiv 2\Omega(0)/(1+z_1),
\label{d}
\end{equation}
\begin{equation}
e \equiv 2-2\Omega(0)-2\lambda(0),
\label{e}
\end{equation}
\begin{equation}
f \equiv \frac{3}{2}\bar{\alpha}\Omega(0)/(1+z_1),
\label{f}
\end{equation}
\begin{equation}
g \equiv \Omega(0)/(1+z_1)^3,
\label{g}
\end{equation}
\begin{equation}
h \equiv \lambda(0),
\label{h}
\end{equation}
\begin{equation}
i \equiv \left[\Omega(0)+\lambda(0)-1\right]/(1+z_1)^2,
\label{i}
\end{equation}
and
\begin{equation}
j \equiv (1+z_1)^{-1}\left\{\Omega(0)(1+z_1)^3+\lambda(0)-\left[\Omega(0)+\lambda(0)-1\right](1+z_1)^2\right\}^{-1/2}.
\label{j}
\end{equation}
Then the general two-component equation to solve is
\begin{equation}
\frac{d^2D}{dz^2}a+\left[3+\frac{b(1+z)-c(1+z)^{-2}}{d(1+z)+e+c(1+z)^{-2}}\right]a(1+z)^{-1}\frac{dD}{dz}+\frac{f(1+z)}{g(1+z)^3+h-i(1+z)^2}D = 0,
\label{14new}
\end{equation}
with boundary conditions
\begin{equation}
D(z_1,z_1)=0,
\label{D14}
\end{equation}
and
\begin{equation}
\frac{dD(z_1,z)}{dz}\mid _{z=z_1} = j.
\label{D15}
\end{equation}

Figure~\ref{dyeroeder1} shows the result of solving
equation~\ref{14new}, with $\bar{\alpha}=1$, in the different
cosmologies, for a source redshift of $z_s = 3.6$, and
Figure~\ref{dyeroeder2} shows the values of $r_dr_{ds}/r_s$, also for
$\bar{\alpha}= 1$. We have tabulated the ratios
$R(\bar{\alpha}=1)/R(\bar{\alpha}=0)$ for the different cosmologies in
Table~\ref{alphaR}. 


\end{document}